\documentclass[aps,prb,twocolumn,psfig,longbibliography,superscriptaddress]{revtex4-2}
\usepackage{graphicx,amssymb,soul,color,amsmath,bm}
\usepackage{epstopdf,hyperref}
\usepackage{braket,dsfont,nicefrac}
\usepackage[mathcal]{euscript}
\usepackage[normalem]{ulem}
\usepackage[utf8]{inputenc}
\usepackage{mathtools,amssymb,lipsum}
\usepackage{multirow}

\usepackage{tikz}
\usetikzlibrary{shapes.geometric}
\usepackage[version=3]{mhchem}
\pgfdeclarelayer{background}
\pgfdeclarelayer{foreground}
\usepackage{appendix}

\hypersetup{colorlinks=true,citecolor=blue,linkcolor=magenta}

\sethlcolor{yellow}
\allowdisplaybreaks

\begin{document}

\title{Recovery of a Luther-Emery phase in the three-band Hubbard model with longer-range hopping}

\author{Luhang Yang}
\email{luhyang@stanford.edu}
\affiliation{Stanford Institute for Materials and Energy Sciences, SLAC National Accelerator Laboratory, 2575 Sand Hill Road, Menlo Park, CA 94025, USA}
\affiliation{Department of Physics, Northeastern University, Boston, MA 02115, USA}

\author{Thomas P. Devereaux}
\affiliation{Stanford Institute for Materials and Energy Sciences, SLAC National Accelerator Laboratory, 2575 Sand Hill Road, Menlo Park, CA 94025, USA}
\affiliation{Department of Materials Science and Engineering, Stanford University, Stanford, CA 94305, USA}

\author{Hong-Chen Jiang}
\affiliation{Stanford Institute for Materials and Energy Sciences, SLAC National Accelerator Laboratory, 2575 Sand Hill Road, Menlo Park, CA 94025, USA}

\begin{abstract}

A lightly doped single-band Hubbard model on a two leg ladder exhibits a Luther-Emery phase, while the three-band Hubbard ladder behaves as a Luttinger liquid upon hole doping. In order to understand this discrepancy, 
we present a systematic density-matrix renormalization group study of the three-band Hubbard model on two-leg cylinders with further-neighbor particle hoppings. The inclusion of the longer-range hopping is motivated by the studies of the single-band Hubbard model in which the further-neighbor hopping terms are suggested to be crucial for the unconventional superconductivity.
When the longer-range hopping parameters are small, the ground state is a Luttinger liquid having mutually commensurate superconducting, charge and spin density wave correlations. Increasing the longer-range hopping drives a transition into a Luther-Emery phase with quasi-long ranged superconducting and charge orders but short-ranged spin-spin correlations. By down-folding the three-band Hubbard model into an effective $t$-$t'$-$J$-$J'$ model, we find that in the Luther-Emery phase, both the nearest and second neighbor kinetic energies are enhanced due to an effective increase of copper-oxygen hybridization. Amplifying inter-cell oxygen orbital hopping mirrors the benefits of reducing the charge transfer energy, causing doped holes to favor oxygen orbitals and strengthening superconducting pairing.

\end{abstract}
\maketitle

\section{Introduction}
\label{sec:intro}

The three-band Hubbard model, which is first proposed by Emery \cite{Emery1987,Emery1988}, can depict the lattice structure of copper and oxygen orbitals in the cuprate superconductors. This model is of particular interest because 1) it takes into account the charge-transfer energy and describes the structure of cuprate materials better than the single-band Hubbard model, and 2) it provides geometrically decoupled spin and charge degrees of freedom in some parameter range. These characters make it a potentially suitable candidate that favors the pairing order. Besides, recent studies on the three-band Hubbard ladder suggest the presence of a pair density wave (PDW) ground state 
\cite{Jiang2023,jiang2023_2}, which is a rare find in the microscopic realization of PDW \cite{Berg2010,Almeida2010,Jaefari2012,Zegrodnik2018,Xu2019,May2020,Zhang2022}.

In the context of high temperature superconductivity, extensive studies have been conducted on the single-band Hubbard model \cite{Arovas2022,Qin2022}. There are some deep relations between the single-band and three-band Hubbard models: the similarity in the fundamental excitations is illustrated by the Zhang-Rice singlet picture \cite{zhangrice}; under certain circumstances, the three-band Hubbard model can be down-folded to an effective single-band Hubbard model or $t$-$J$ model \cite{jiang2023singleband,ESKES1989424,Eskes1991,Eskes1991_2}, where similar ground state properties 
have been revealed in recent decades \cite{Raimondi1996_1,Raimondi1996_2,Belinicher1993,jiang2023singleband,Huang2017,Huang2018}. The ground state of the single-band Hubbard model at half filling on a two dimensional square lattice is a Mott insulator with long range magnetic order \cite{Hirsch1985,Yokoma2006}. The three-band Hubbard model at half filling also has been found to exhibit AFM order \cite{Dopf1990,Yanagisawa2001}.

However, discrepancies between these two models are not negligible. 
Although it has been extensively studied in the context of high temperature superconductivity, no evidence of PDW has been found in the single-band Hubbard model. More strikingly, the single-band Hubbard ladder exhibits a Luther-Emery phase with dominant superconducting and charge orders upon light doping \cite{ShenYang2023,Hongchen2018}, while the three-band Hubbard ladder is a Luttinger liquid for the same doping concentrations \cite{Jiang2023,Song2021}.

In order to understand these discrepancies, we study 
the three-band Hubbard model with longer-range hopping on a two-leg ladder. The inclusion of the longer-range hopping is motivated by previous studies on the single-band Hubbard model, in which the further neighbor particle hoppings are essential 
for superconductivity \cite{Aichhorn2006,Ponsioen2019,Jiang2019Hub,Arovas2022,Qin2022,Yifan2020,Shengtao2021,Jiang2023Six}. 
We study the ground state properties of the lightly doped three-band Hubbard model on two-leg cylinders of Lieb lattices. The further neighbor hopping terms we introduce here are the hopping between copper sites in the adjacent cells with coefficients $t_{dd}$, and the hopping between oxygen sites in the adjacent cells with coefficients $t_{pp}^{\prime}$. By tuning the hopping parameters, we get a phase diagram with a Luttinger Liquid phase with intertwined PDW, charge density wave (CDW) and spin density wave (SDW) correlations 
when both hoppings are close to zero, and a Luther-Emery SC phase with $d$-wave symmetry when these two parameters are greater than a critical value.

The rest of the manuscript is organized as follow: in Sec. \ref{sec:model_phase_diagram} we describe the model Hamiltonian and illustrate the phase diagram; in Sec. \ref{sec:gs_correlations} we show various ground state correlations and discuss how the hopping parameters affect the ground state properties; in Sec. \ref{downfolding} we present an analysis of down-folding the three-band Hubbard model to an effective $t$-$t^{\prime}$-$J$-$J^{\prime}$ model using the exact diagonalization (ED) method; and finally in Sec. \ref{sec:conclusion} we summarize our results.

\section{Model and phase diagram}
\label{sec:model_phase_diagram}
\begin{figure}
	\centering
  \includegraphics[width=0.45\textwidth]{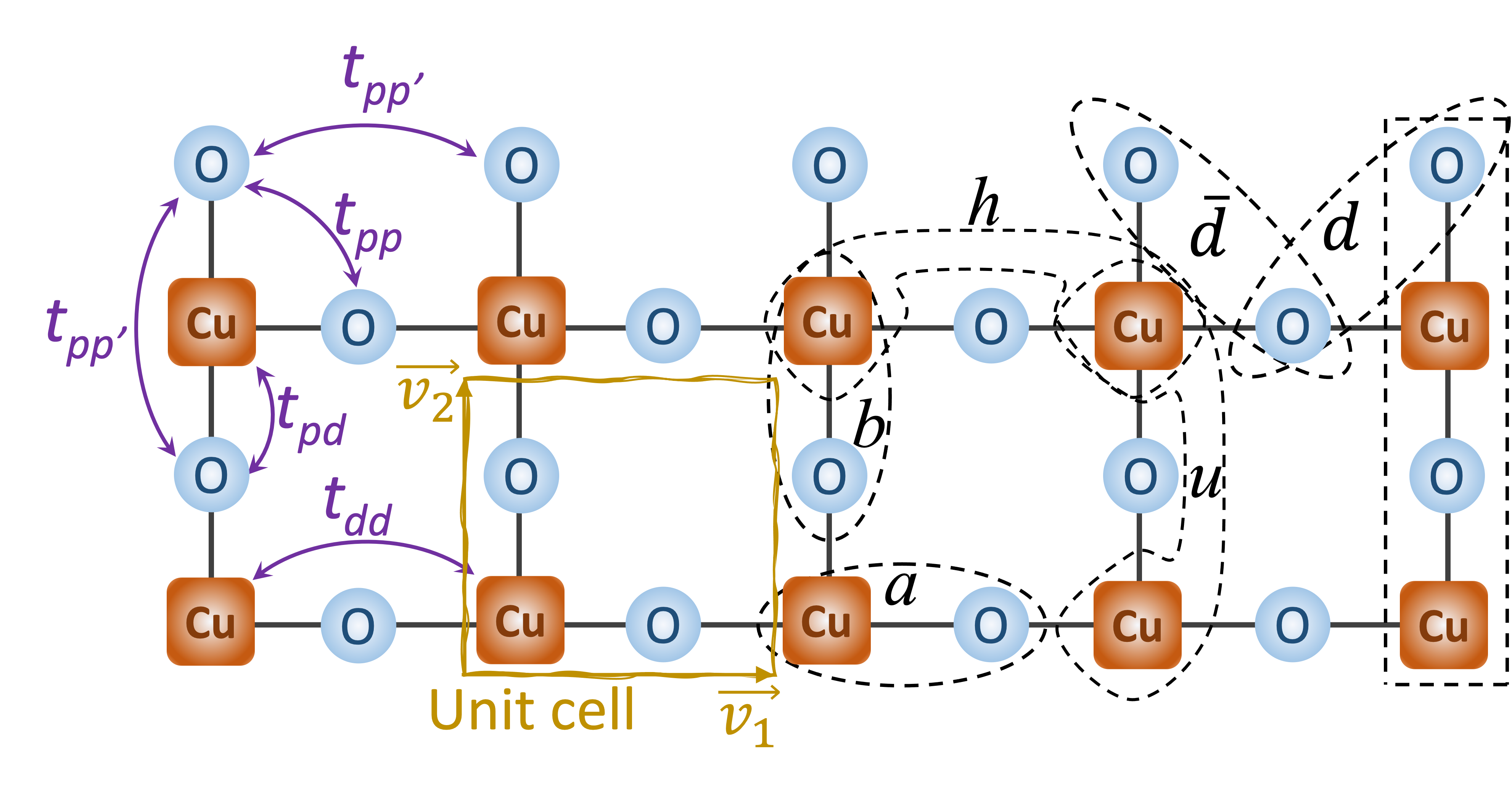} \hfill
  \includegraphics[width=0.45\textwidth]{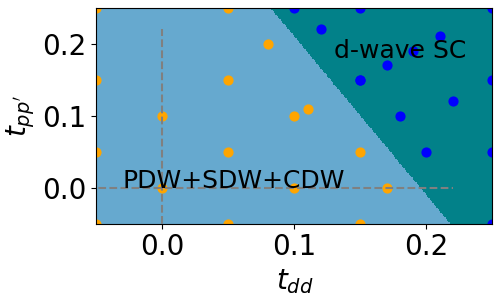}
	\caption{Top: The three-band Hubbard model on a Lieb lattice. The hopping parameters are shown on the lattice. Several types of copper-copper bonds are marked in dashed lines. We add an additional column on the right end to facilitate convergence. Here $\vec{v}_1$ and $\vec{v}_2$ are basis vectors in the unit cell. Bottom: Phase diagram as a function of $t_{dd}$ and $t_{pp^\prime}$. The region in orange represents the PDW+CDW+SDW phase, and the region in blue is the SC phase.
 } \label{fig:phase_diagram}
\end{figure}

We use the density-matrix renormalization group (DMRG) method \cite{White1992,White1993,Ostlund1995} based on iTensor library~\cite{Fishman2022,Fishman2022_2} to study the ground state phase diagram of the three-band Hubbard model with further neighbor hoppings on two-leg ($L_y=2$) Lieb lattice cylinders with
length up to $L_x=64$. By keeping bond dimensions up to $m=7000$ we make sure the truncation error remains in the order of $10^{-7}$ or below.
The Lieb lattice we simulate is shown in the top panel of Fig. \ref{fig:phase_diagram}. The ``squares'' and ``circles'' represent copper and oxygen orbitals, respectively. We implement periodic (open) boundary condition along the vertical (horizontal) direction.

We study the following model Hamiltonian in the hole language: 
\begin{eqnarray}
H&=&H_{tb}+H_{int} \\
    H_{tb} & = & -T^{pd} - T^{pp} -T^{dd} - T^{pp'}\\ \nonumber
    & + & \Delta_{pd}\sum_{i}n_{i}^p \\
    H_{int} &=& U_d\sum_{i}n_{\uparrow,i}^d n_{\downarrow,i}^d + U_p\sum_{i}n_{\uparrow,i}^p n_{\downarrow,i}^p,
\label{hami}
\end{eqnarray}
where the kinetic term is defined as $T^{\alpha\beta} = \sum_{i,j;\sigma}t_{\alpha\beta}(c_{ i,\sigma}^{\alpha \dagger} c_{j,\sigma}^{\beta} + h.c.)$, and $\alpha, \beta$ belongs to each kind of orbitals. We take into account four types of hopping terms in this model as shown in the top panel of Fig. \ref{fig:phase_diagram}: 1) $t_{pd}$ between the adjacent oxygen and copper sites; 2) $t_{pp}$ between the nearest $p_x$ and $p_y$ orbitals; 3) $t_{dd}$ between copper sites in the adjacent unit cells; 4) $t_{pp'}$ between the same type of oxygen sites in the adjacent cells.
In this work, we keep $U_d = 8$, $U_p=3$, $t_{pd}=1$, $t_{pp}=0.5$, $\Delta_{pd}=3$ \cite{Hybertsen1989,Nishimoto2002,White2015} unless otherwise specified. $t_{dd}$ and $t_{pp}^{\prime}$ are the tuning parameters. We follow the sign convention in Ref. \cite{Jiang2023},
which is equivalent to the original Emery model by a gauge transformation.

We calculate the following correlation functions 
to study the ground state properties of the systems. These include the spin-spin correlation function
\begin{equation}
S(r) = \langle S^z_0S^z_r\rangle, 
\label{Sr}
\end{equation}
the charge density-density fluctuation correlation function 
\begin{equation}
D(r) = \langle n_0n_r\rangle - \langle n_0\rangle \langle n_r\rangle,
\label{Nr}
\end{equation}
the single particle Green function 
\begin{equation}
G(r) = \langle c^{\dagger}_{\uparrow,0}c_{\uparrow,r}\rangle, 
\label{Sr}
\end{equation}
and the spin-singlet SC pair-pair correlation function
\begin{equation}
\Phi(r) = \langle \Tilde{\Delta}^\dagger_0\Tilde{\Delta}_r\rangle .
\label{Pr}
\end{equation}
Here $\Tilde{\Delta}^\dagger$ represents the spin-singlet Cooper pair creation operator
between neighboring sites 
\begin{equation}
\Tilde{\Delta}^\dagger_i = \frac{1}{\sqrt{2}}(c^\dagger_{i,\downarrow}c^\dagger_{i+1,\uparrow} - c^\dagger_{i,\uparrow}c^\dagger_{i+1,\downarrow}).
\end{equation}

In the bottom panel of Fig.\ref{fig:phase_diagram} we show the phase diagram as a function of $t_{dd}$ and $t_{pp}^\prime$ at $\delta = 1/8$  hole doping concentration. 
When both
hopping parameters are around zero, the system is in a Luttinger liquid phase with intertwined PDW, CDW and SDW correlations.
In order to find the strongest signal of the pairing order, we have computed the SC pairing correlations on different bonds (see Appendix \ref{app:pairing_bonds}) and find that $\Phi_{hh}(r)$,
i.e., the spin-singlet SC pairs on nearest copper sites along x-direction, is the dominant SC component.
By increasing both hopping coefficients to the positive values, the
spin-singlet pairing correlations
are further enhanced
and the system
eventually undergoes a quantum phase transition to a $d$-wave SC phase. The properties of this $d$-wave SC phase is consistent with that of a Luther-Emery liquid state with quasi-long-range SC and CDW correlations but short-range spin-spin correlation.
Different with the Luttinger liquid phase, the spin-singlet pairing correlation 
between adjacent copper sites along y-direction $\Phi_{uu}(r)$ becomes dominant over all the other bonds here, and exhibits a $d$-wave symmetry.

\section{Ground state correlations}
\label{sec:gs_correlations}

When $t_{dd}$ and $t_{pp}^{\prime}$ are both small and positive, the ground state of the system is consistent with that of the Luttinger liquid phase with power-law single particle,
PDW, SDW and CDW correlations.
However, different with the single-band Hubbard model, the SC correlation
(Fig. \ref{fig:corr_LL_pair}(b)) in this case has a spatial oscillation with sign change, which can be described by the following formula:
\begin{equation}
    \langle \Tilde{\Delta}_0^{\dagger} \Tilde{\Delta}_r \rangle = \frac{A\cdot cos(\omega r+\phi)}{r^{\kappa_{pdw}}} + \frac{B}{r^{\kappa_{sc}}}
\end{equation}

If the value of $B\leq A$,
the overall fluctuation has a spatial oscillation around zero, which suggests the presence of the PDW order.
Our results are consistent with this, where we find that $A\approx 0.01, B\approx 0.001$ when $t_{dd}=t_{pp}^{\prime}=0$. For $t_{dd}=t_{pp}^{\prime}=0.2$, we find that $A\approx 0.006$, $B\approx 0.02$.

The charge density properties of the system can be described by the charge density profile  $n_\alpha(x,y)$ and its rung average $\rho_\alpha(x)=\sum_{y=1}^{L_y}n_\alpha(x,y)/L_y$. Consistent with previous studies \cite{Jiang2023,jiang2023_2}, the spatial decay of the CDW correlation at long distance is dominated by a power-law with an exponent $\kappa_c$ with two ordering wavevectors $Q=2\pi\delta$ and $2Q$. The value of the exponent $\kappa_c$ can be obtained by fitting the charge density oscillations $\rho(x)$ with a generalized Friedel oscillation formula induced by the open boundaries of the cylinder \cite{White2002Friedel,Jiang2023}
\begin{equation}
    \rho(x) = \frac{A_Q\ast {\rm cos}(Q x + \phi_1)}{x^{\kappa_c/2}} +  \frac{A_{2Q}\ast {\rm cos}(2Q x + \phi_2)}{x^{\kappa_c/2}} +n_0,
\label{n_profile}
\end{equation}
where $A_Q$ and $A_{2Q}$ are the amplitudes, $\phi_1$ and $\phi_2$ are the phase shifts and $n_0$ is the mean density.
It has been shown that 
the $2Q$ charge order usually competes with the superconductivity \cite{Arovas2022,Agterberg2015}. 
which is also
consistent with our results. We will see that when the system enters the superconducting phase, the $2Q$ mode is suppressed.

\begin{figure}
\centering
\includegraphics[width=.45\textwidth]{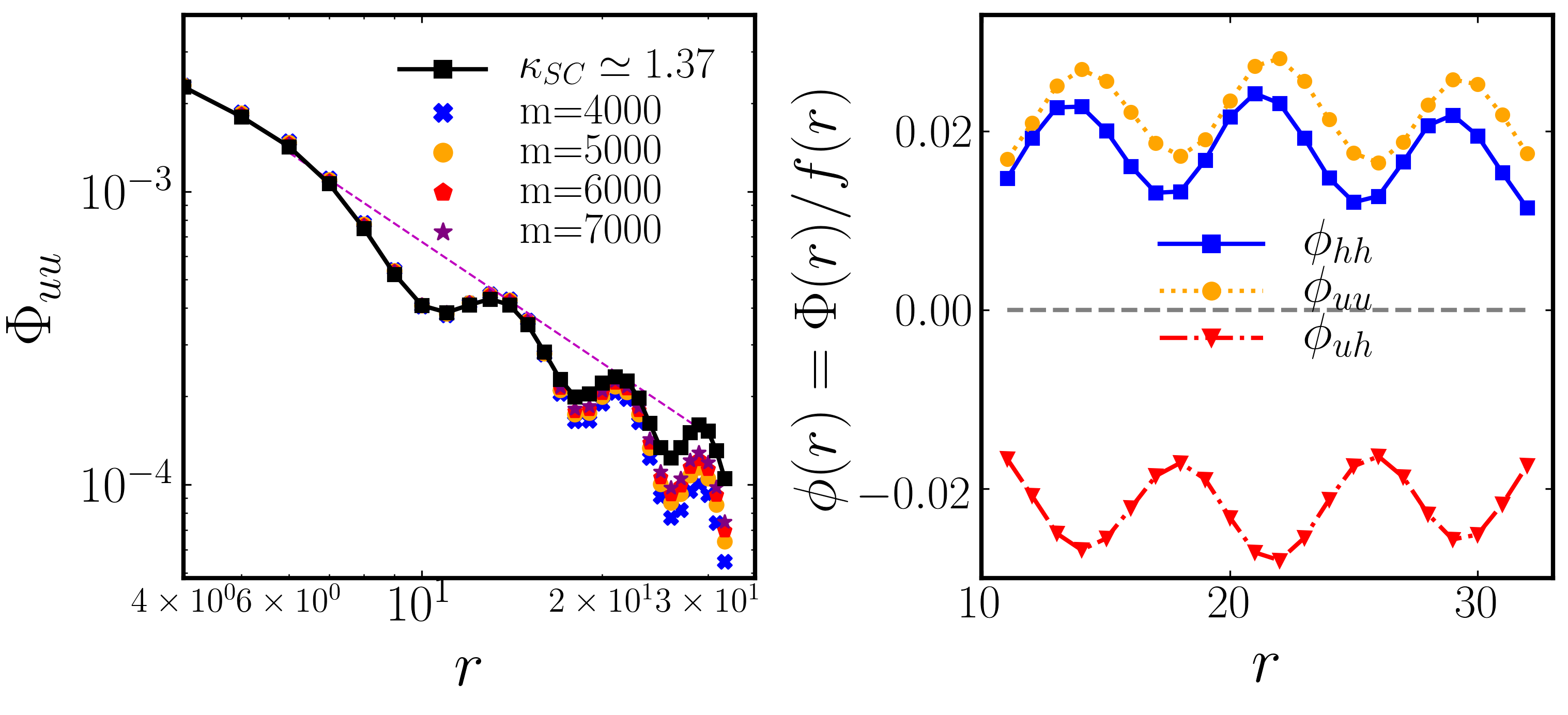}
\caption{Correlations for $t_{dd}=t_{pp}^{\prime}=0.2$. Left: Singlet Pairing correlations on $u$ bonds in log scale for different number of states $m$. The black curve with square symbols represents the result of a $m\rightarrow \infty$ extrapolation; Right: the normalized pairing correlations on $u-u, h-h,$ and $u-h$ bonds. $f(r)$ is the envelope function. }
\label{fig:corr_SC_pair}
\end{figure}

\begin{figure}
\centering
\includegraphics[width=.45\textwidth]{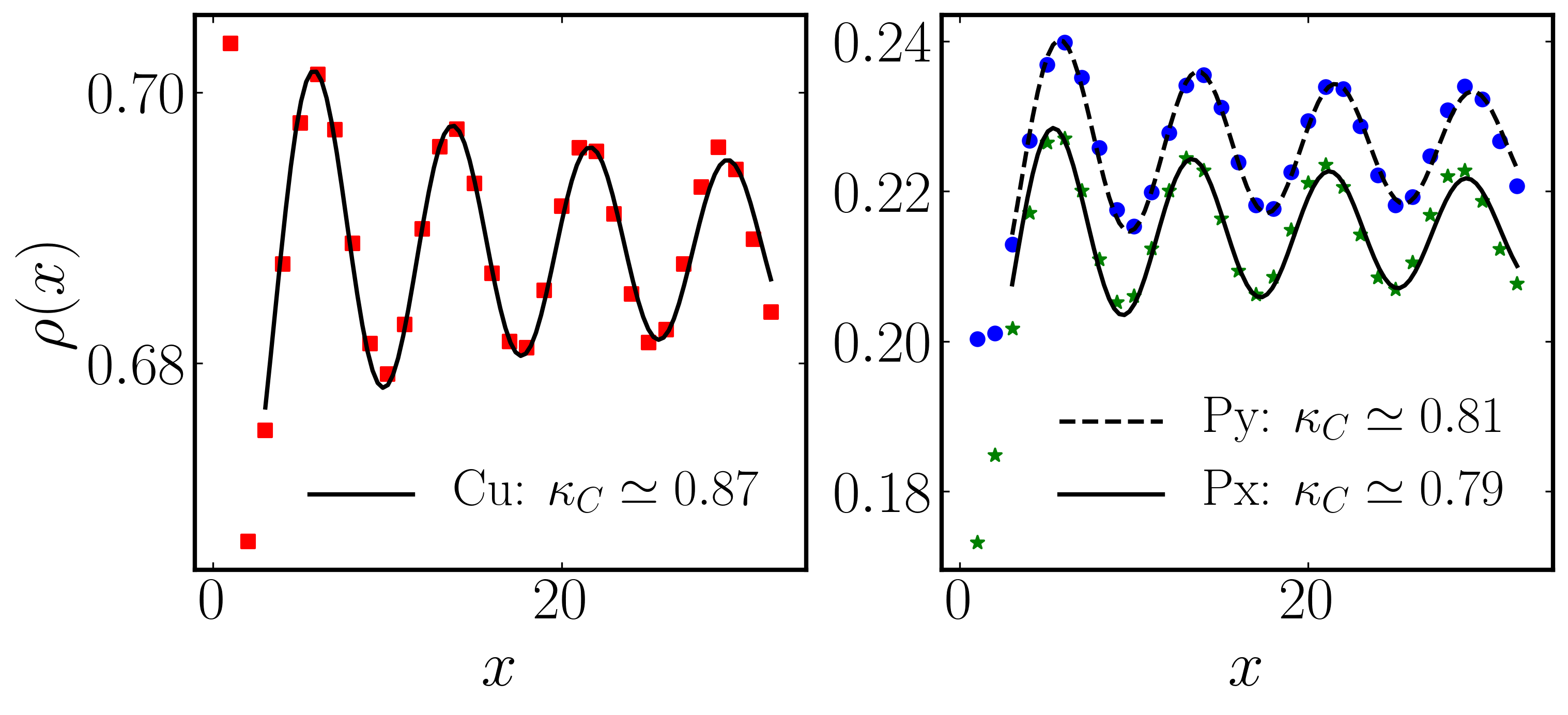}
\caption{Local density profiles for $t_{dd}=t_{pp}^{\prime}=0.2$. Left: density distribution on copper orbital; Right: density distribution on oxygen orbital.  $\kappa_C$ is the scaling exponent obtained by curve fitting.}
\label{fig:corr_SC_local}
\end{figure}

By increasing
the parameters to the larger positive values,
the system enters a distinct $d$-wave SC phase with dominant power-law superconducting correlations but exponentially decaying single-particle and spin-spin correlations.
In the $d$-wave SC phase, the short-ranged spin-spin correlation
is mutually commensurate with the charge 
correlation at the wavevector $Q$, although both of
them are incommensurate on the finite lattice with open boundaries in the longer direction.

Contrary to the Luttinger liquid phase with small further neightor electron hoppings, we find that in this $d$-wave SC phase the coefficient $B$ is greater than $A$ . Moreover, the charge density modulation has only one characteristic wave vector $Q$ (Fig. \ref{fig:corr_SC_local}). The $2Q$ mode is suppressed while the SC order is enhanced. The symmetry of the SC correlations can be determined by comparing the relative signs of the SC pair-pair correlations between different bonds. The right panel of Fig. \ref{fig:corr_SC_pair} shows that the pairing correlations between $u$ bonds display values in opposite sign with the correlations between $u$ and $h$ bonds (see in Fig. \ref{fig:phase_diagram} for the definition of the bonds). Besides, the multiplication of $\kappa_c$ and $\kappa_{sc}$
is close to 1. All these suggest that the system is in a Luther-Emery phase, with a $d$-wave pairing symmetry.

The results above show that the inter-cell hopping terms 
enhance the SC correlations while suppressing the spin-spin correlation.
This is closely connected to the results of charge transfer energy which also suggests that the density distribution is associated with the intertwined orders in the charge transfer insulators \cite{Scalettar1991}. In order to understand how these two pictures reconcile, we study how the hopping terms affect the density distributions on each orbital.
It is shown in \cite{Nishimoto2002,White2015} that for the undoped three-band Hubbard model, about $70\%$ of the holes
are on the copper sites.  
Upon (hole) doping, however, most of the doped holes will occupy the oxygen sites \cite{Scalettar1991}.
Interestingly, if we turn on the further neighbor electron hoppings $t_{dd}$ and $t_{pp}^{\prime}$, we find that the effect of increasing $t_{pp}^{\prime}$ is equivalent to decreasing the effective $\Delta_{pd}$, where the average copper density will decrease (See Appendix \ref{app:charge_transfer}). On the contrary, the influence of $t_{dd}$ on density distribution is negligible.

\begin{figure}
	\centering
	\includegraphics[width=0.45\textwidth]{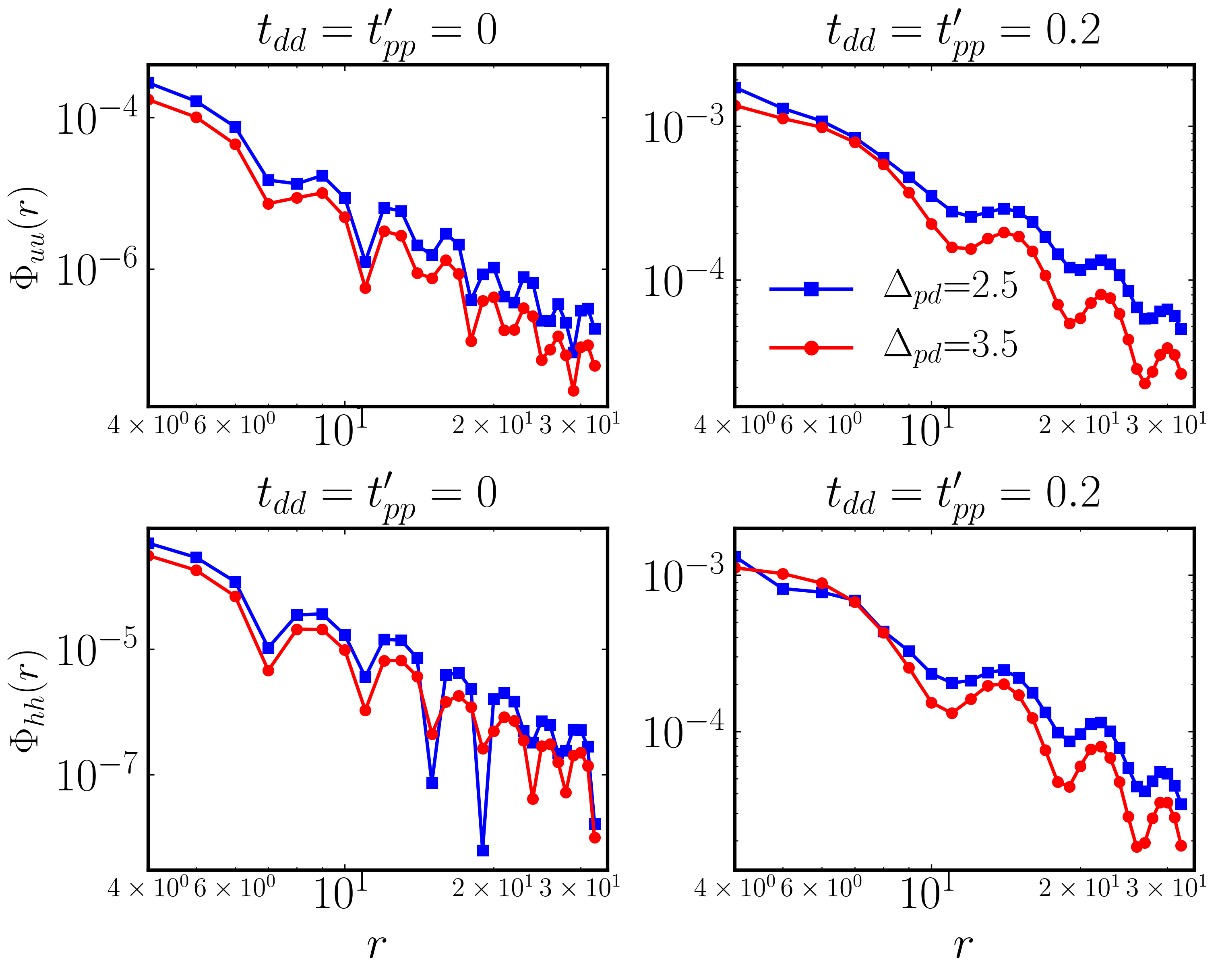}
    \caption{Comparison of singlet pairing correlations with different values of charge transfer energy in each of the PDW+CDW+SDW and SC phases.}
	\label{fig:change_deltapd}
\end{figure}

As a complementary comparison, we have computed the SC pairing correlation for different charge transfer energies $\Delta_{pd}$ while all the other parameters remain the same. In Fig. \ref{fig:change_deltapd} we show the SC pairing correlations for two representative cases in each of the two phases,
where we find that for both phases, decreasing $\Delta_{pd}$ can enhance the SC pairing correlations.

\section{Down-folding the three-band Hubbard model to an effective $t$-$t^{\prime}$-$J$-$J^{\prime}$ model}
\label{downfolding}

The low energy physics of the three-band Hubbard model (or CuO$_2$ plane) can be mapped to an effective $t$-$t^{\prime}$-$J$-$J^{\prime}$ model in Eq. \ref{t1t2}. Following the prescription in Ref. \cite{ESKES1989424,Eskes1991,Eskes1991_2}, we present the down-folding for the three-band Hubbard model with further neighbor hopping terms. In this scheme we consider two types of small clusters: the Cu$_2$O$_7$ cluster with two unit cells aligned, and  the Cu$_2$O$_8$ cluster with the two unit cells along the diagonal direction. Specifically, the hopping parameter $t$ ($t^{\prime}$) is determine by the spin singlet and triplet energy splitting of a Cu$_2$O$_7$ (Cu$_2$O$_8$) cluster; $J$ ($J^{\prime}$) is determined by the energy difference between bonding and anti-bonding states of a Cu$_2$O$_7$ (Cu$_2$O$_8$) cluster. The results are shown in Table \ref{table1}. When the three-band Hubbard model is in the $d$-wave SC phase, we find in the effective model that the hoppings to both the nearest ($t$) and next nearest neighbors ($t^{\prime}$) are enhanced, as well as the spin exchange coupling $J$. The strengthening of the hybridization between orbitals 
makes the pairing orders more favorable. 

\begin{table}[]
\centering
\begin{tabular}{ |p{1cm}|p{1cm}||p{1cm}|p{1cm}|p{1cm}|p{1cm}|  }
 \hline
 \multicolumn{2}{|c|}{\ } &\multicolumn{2}{|c|}{$\rm Cu_2O_7$} & \multicolumn{2}{|c|}{$\rm Cu_2O_8$}\\
 \hline
 $t_{dd}$ & $t_{pp}^{\prime}$ & $t$ &$J$ & $t^{\prime}$ & $J^{\prime}$ \\
 \hline
 0  & 0    &0.34  &0.17 &-0.14 &0.01\\
 0  & 0.25 &0.45  &0.27 &-0.21 &0.04\\

 0.25 & 0    &0.41  &0.34 &-0.14 &0.01\\
 0.25 & 0.25 &0.52  &0.47 &-0.21 &0.04\\
 \hline
\end{tabular}
\caption{\label{table1} Coefficients of the effective $t$-$t^{\prime}$-$J$-$J^{\prime}$ model. There is an overall minus sign before each hopping coefficient.}
\end{table}
\qquad

\begin{eqnarray}
    H_{t-t^{\prime}-J-J^{\prime}} 
\label{t1t2}
    & = & -
    t\sum_{\langle i,j \rangle \sigma}(c_{i,\sigma}^\dagger c_{j,\sigma} + h.c.) \\ \nonumber
    & - & t^{\prime}\sum_{\langle\langle i,j \rangle\rangle\sigma}(c_{i,\sigma}^\dagger c_{j,\sigma} + h.c.) \\ \nonumber
    & + & J\sum_{\langle i,j \rangle \sigma}(\vec{S}_i\cdot \vec{S}_{j} - \cfrac{1}{4} n_in_{j}) \\ \nonumber
    & + & J^{\prime}\sum_{\langle\langle i,j \rangle\rangle}(\vec{S}_i\cdot \vec{S}_{j} - \cfrac{1}{4} n_in_{j}) \\ \nonumber
\end{eqnarray}

\begin{figure}
\centering
\includegraphics[width=.45\textwidth]{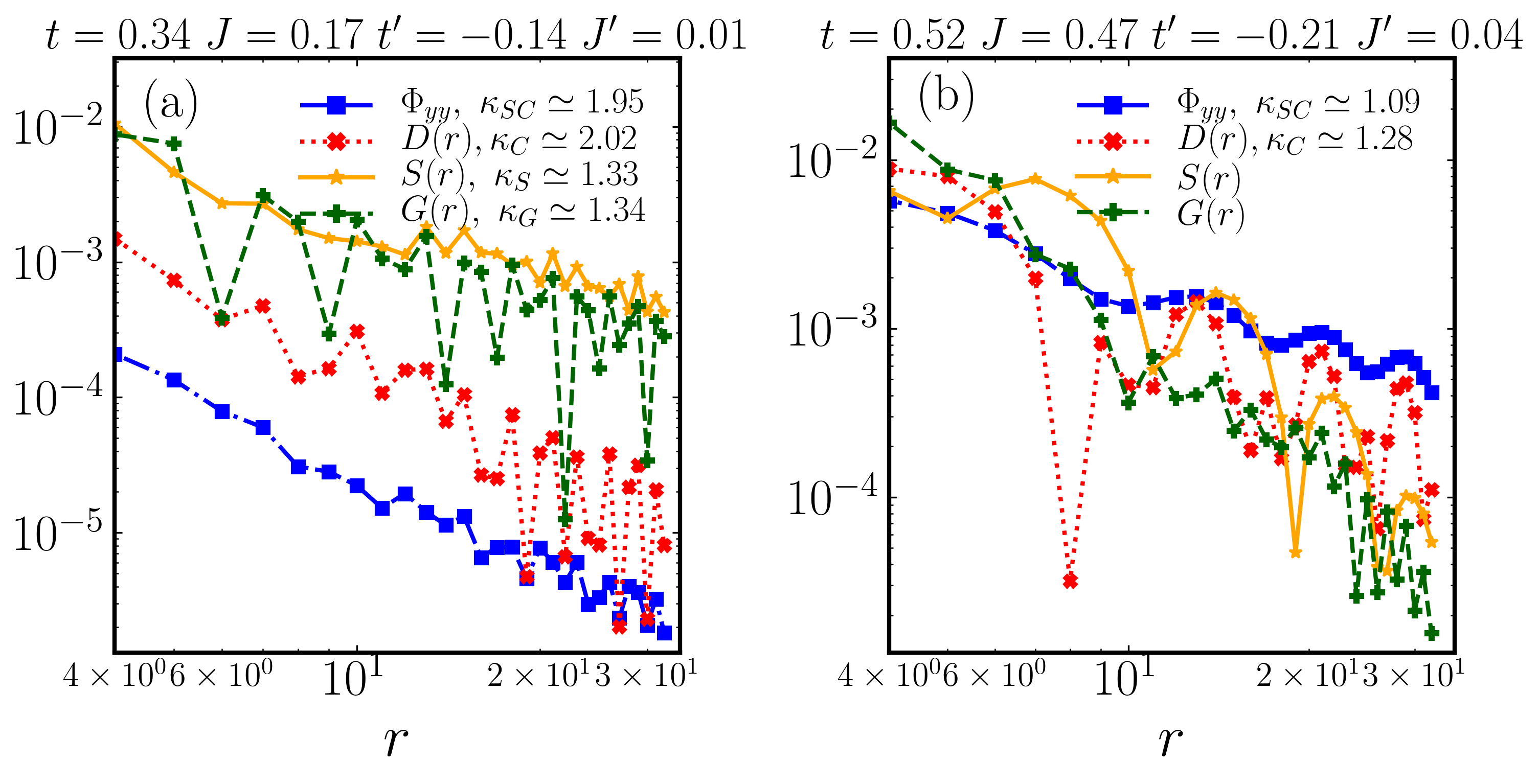}
\caption{Ground state correlations of the $t$-$t^{\prime}$-$J$-$J^{\prime}$ model. (a): The results for $t=0.34$, $t^{\prime}=-0.14$, $J=0.17$, $J^{\prime}=0.01$. The corresponding three-band Hubbard model is in the LL phase. (b): The results for $t=0.52$, $t^{\prime}=-0.21$, $J=0.47$, $J^{\prime}=0.04$. The corresponding three-band Hubbard model is in the SC phase.}
\label{fig:tt'JJ'SC}
\end{figure}

To further support this argument, we have also calculated the ground state properties of the $t$-$t^{\prime}$-$J$-$J^{\prime}$ model on a $N=64\times 2$ ladder with DMRG, using the parameters listed in Table \ref{table1}. We choose two representative sets of parameters
in the first and last rows of the table to implement the calculations. The corresponding three-band Hubbard models of these two data points are in the Luttinger liquid phase and $d$-wave SC phase respectively. The results are presented in Fig. \ref{fig:tt'JJ'SC}. We can see in the left panel that all the correlations decay as a power-law, with dominant single particle and spin correlations.
However in the right panel,
the SC pairing and CDW correlations become dominant
while the spin and single particle correlations are short-ranged which decay exponentially. Similar with three-band Hubbard model, the pairing symmetry of the SC correlations
is also $d$-wave in this case.

\section{Conclusions}
\label{sec:conclusion}

To summarize, we have studied the ground state properties of the lightly doped three-band Hubbard model with longer-range hopping terms on two-leg cylinders. By tuning these hopping coefficients positively, we observed a quantum phase transition from a Luttinger liquid phase characterized by intertwined PDW, CDW, and SDW correlations to a $d$-wave SC phase (i.e., Luther-Emery phase). Through this transition, the pairing order intensifies, changing from a mixed SC and PDW order to a predominantly $d$-wave symmetry SC order. This transition is underlined by modifications in the band structure and density distributions stemming from the tuning of hopping parameters. Our computational analyses pinpointed an intriguing correlation: the increase in the $t_{pp}^\prime$ parameter mimics the effect of a reduced effective charge transfer energy, $\Delta_{pd}$.

By down-folding the three-band Hubbard model to the effective $t$-$t^{\prime}$-$J$-$J^{\prime}$ model, we observed an increase of both the nearest and second neighbor hopping parameters and the ratio of $J/t$, when entering the Luther-Emery phase. This finding sheds light on the question ``why the lightly doped three-band Hubbard ladder behaves as a Luttinger liquid but not the Luther-Emery liquid, as the single-band Hubbard model does''. From our small cluster study we found that although these two models were closely related, the parameters of the original three-band Hubbard model did not have a strong enough hybridization between orbitals. However, by introducing longer-range hopping terms, the Luther-Emery phase emerges in the vicinity of the Luttinger liquid phase.

In this study, we have examined the effect of further-neighbor particle hopping terms on the ground state properties of the three-band Hubbard model on two-leg cylinders. It is worth noting that the recovery of the Luther-Emery phase on the three-band Hubbard model can also be achieved by introducing long-range Coulomb interactions \cite{jiang2023_2}.
An intriguing avenue for future research would be to ascertain whether this influence persists in wider systems and at higher doping levels. A comprehensive understanding would also necessitate the exploration of the combined effects of further-neighbor hopping and long-range Coulomb interactions, for which previous studies have demonstrated that the later one is crucial for enhancing or in some instances even inducing the PDW order and superconductivity \cite{Chen2021,Qu2022,Peng2023,Jiang2023,jiang2023_2}.

\section{Acknowledgement}

This work was supported by the Department of Energy, Office of Science, Basic Energy Sciences, Materials Sciences and Engineering Division, under Contract DE-AC02-76SF00515. Computational work was performed on the Sherlock cluster at Stanford University and on resources of the National Energy Research Scientific Computing Center, supported by the U.S. DOE, Office of Science, under Contract no. DE-AC02-05CH11231.

\appendix
\renewcommand\thefigure{\thesection.\arabic{figure}}

\section{Effects of hopping parameters and charge transfer energy on density distribution}
\label{app:charge_transfer}
\setcounter{figure}{0} 
In order to demonstrate how the hopping coefficients affect the number density, we compute the average density on copper and oxygen sites for systems with length $Lx=48$ for two cases: (1) half filling and (2) $1/8$ hole doping. Fig. \ref{fig:local_density} shows the average local densities on Cu, O$_x$ and O$_y$ sites with different values of intra-orbital hoppings. 
In both of the half filling and the hole doped regimes, the density on copper sites mainly depends on the values of $t_{pp}^\prime$, and keeps nearly unchanged with $t_{dd}$. When $t_{pp}^\prime$ is increasing,  the occupation number on Cu sites decreases and more holes go to oxygen sites, which is similar with the effect of decreasing the charge transfer energy $\Delta_{pd}$.

Nevertheless, the doped holes (the increase of the hole density upon doping) have both $t_{dd}$ and $t_{pp}^{\prime}$ dependencies. With the increase of $t_{pp}^{\prime}$, the doped holes on both $O_x$ and $O_y$ decreases. With the increase of $t_{dd}$, the doped holes on $O_x$ and $O_y$ have opposite trends, and the number of doped holes on copper sites decreases.

\begin{figure}
	\centering
	\includegraphics[width=0.45\textwidth]{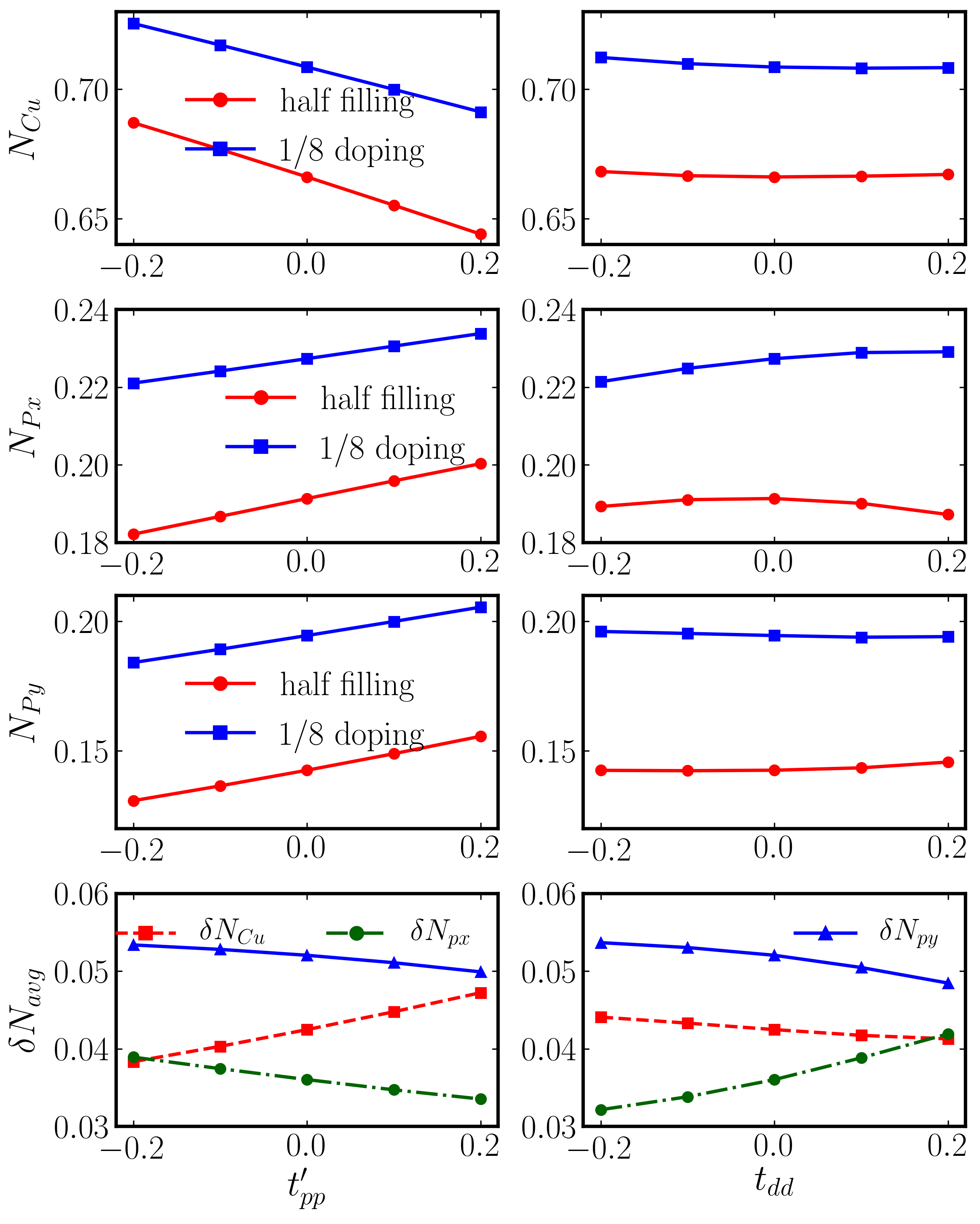}
    \caption{Number density measurements on Cu, O$_x$, and O$_y$ sites.}
	\label{fig:local_density}
\end{figure}

\begin{figure}
	\centering
	\includegraphics[width=0.3\textwidth]{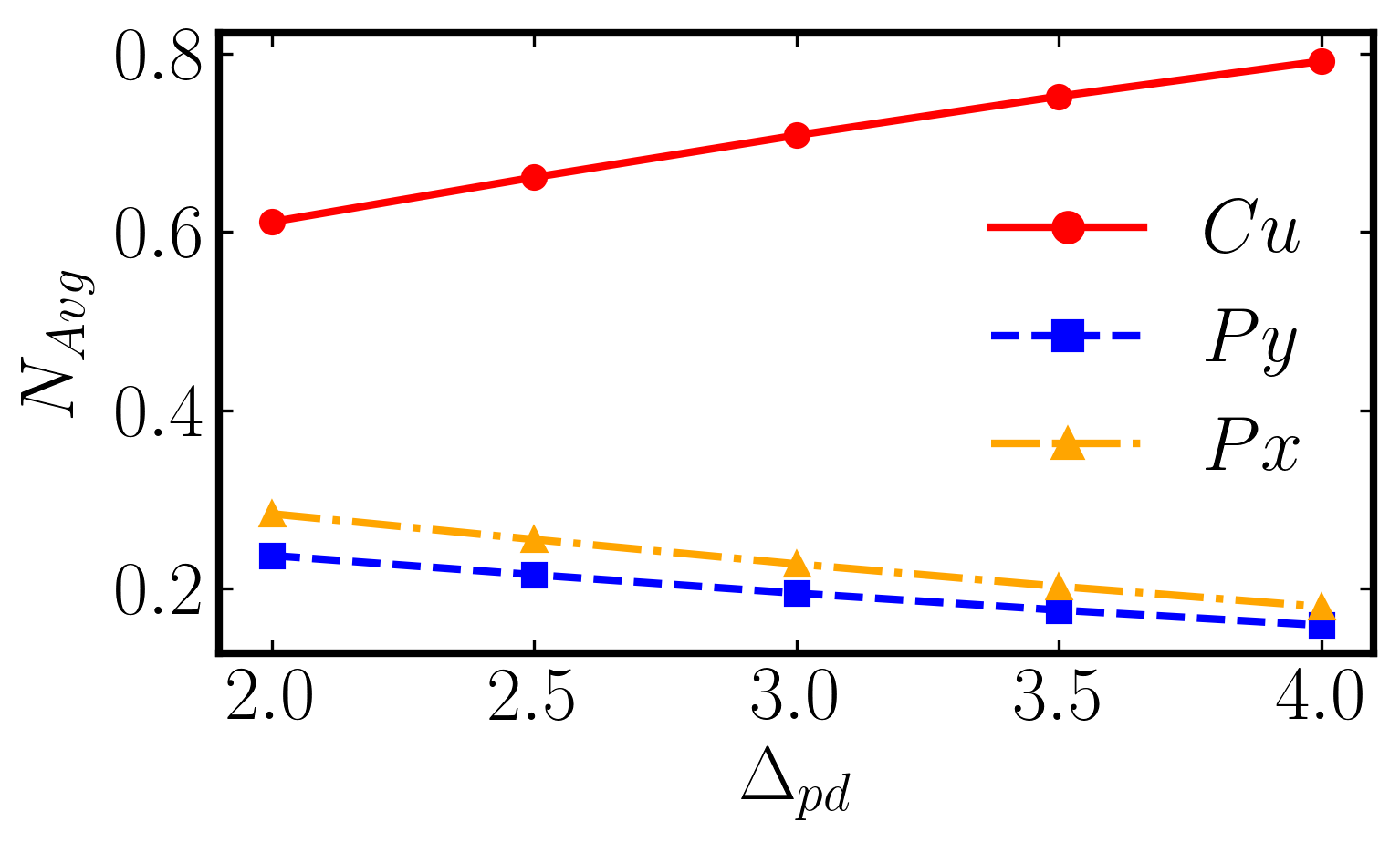}
    \caption{Average number density as a function of $\Delta_{pd}$. Computed by DMRG for system size Nx=48 with $1/8$ doping.}
	\label{fig:local_density_deltapd}
\end{figure}

\section{Comparison of the pairing orders on different bonds}
\label{app:pairing_bonds}
\setcounter{figure}{0} 
In Fig. \ref{fig:pair_all_bonds} we show the pairing correlations on all types of bonds illustrated in the top panel of the schematic Figure \ref{fig:phase_diagram}. When the system is in the PDW phase, the PDW order on the $h$ bonds is slightly stronger than all the others, although all of the pairing orders are quasi-long ranged. However in the $d$-wave SC phase, the SC order on $u$ and $h$ bonds are dominant, and are stronger than the others by a few orders of magnitude.

\begin{figure}
\centering
\includegraphics[width=.22\textwidth]{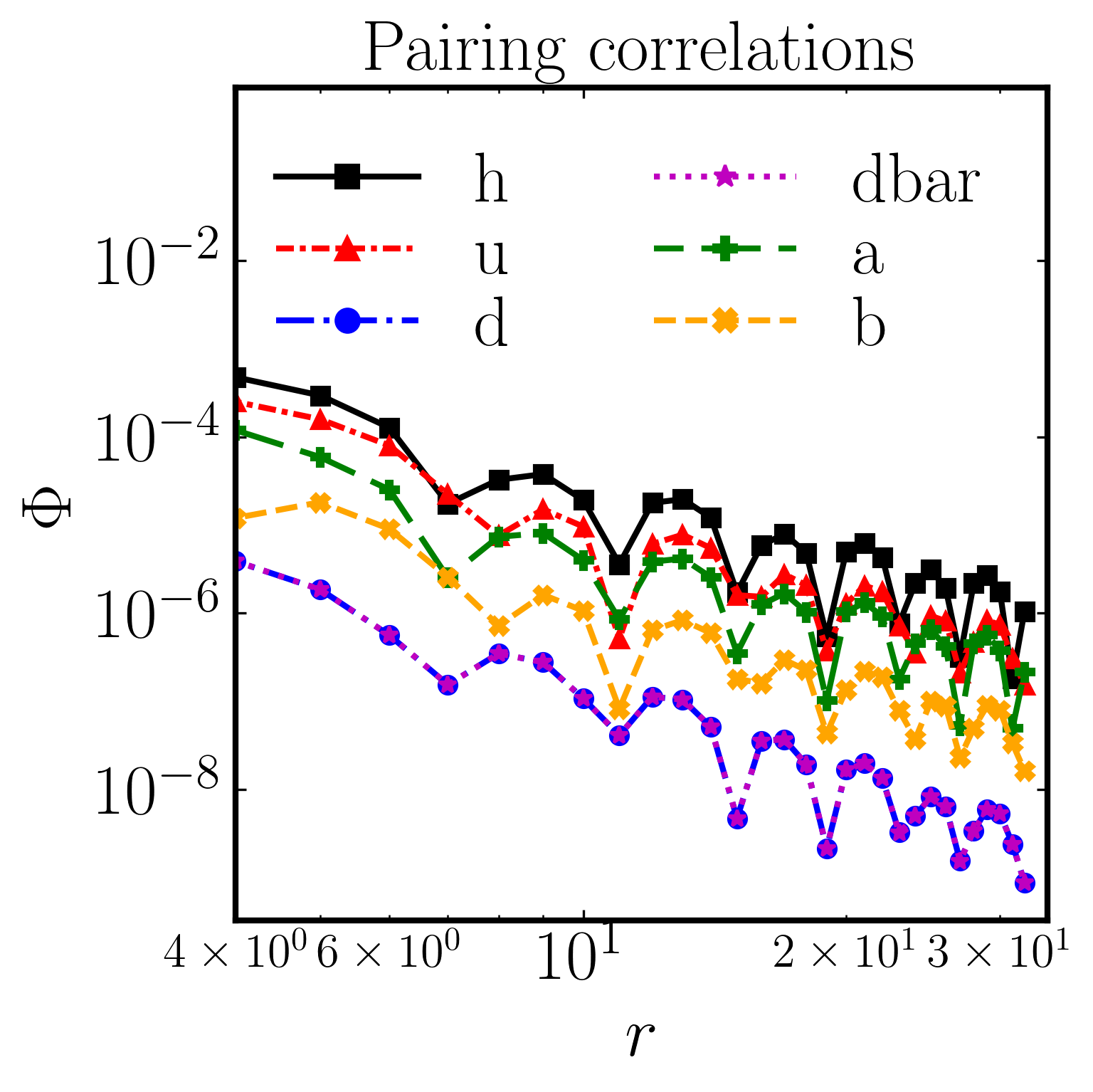}\hfill
\includegraphics[width=.22\textwidth]{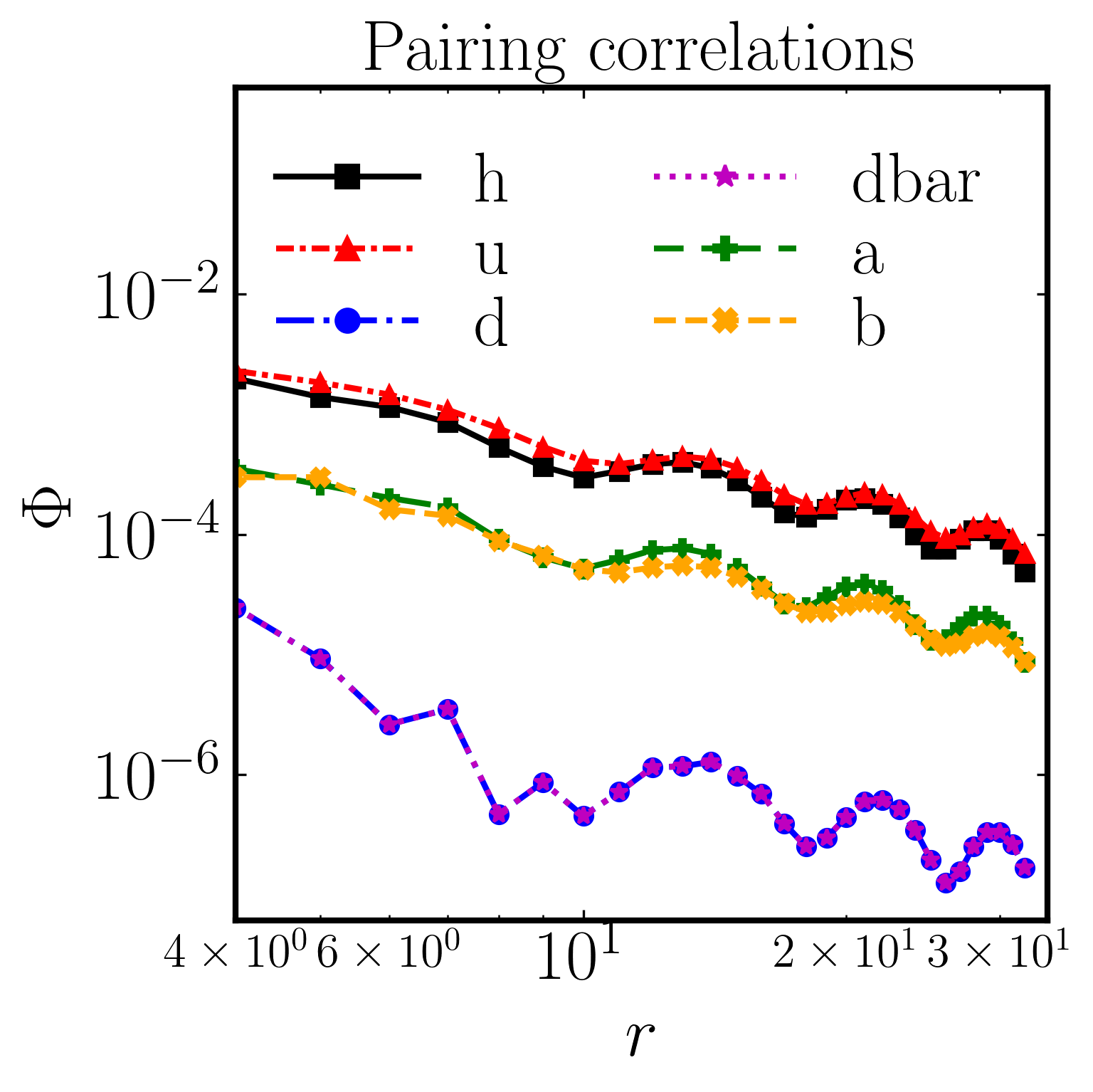}
\caption{Left: Comparison of singlet pairing correlations on different bonds for $t_{dd} = t_{pp}^\prime = 0$. Right: Comparison of singlet pairing correlations on different bonds for $t_{dd} = t_{pp}^\prime = 0.2$.}
\label{fig:pair_all_bonds}
\end{figure}

\section{Ground state correlations in the LL phase}
\setcounter{figure}{0} 

\begin{figure}
\centering
\includegraphics[width=.45\textwidth]{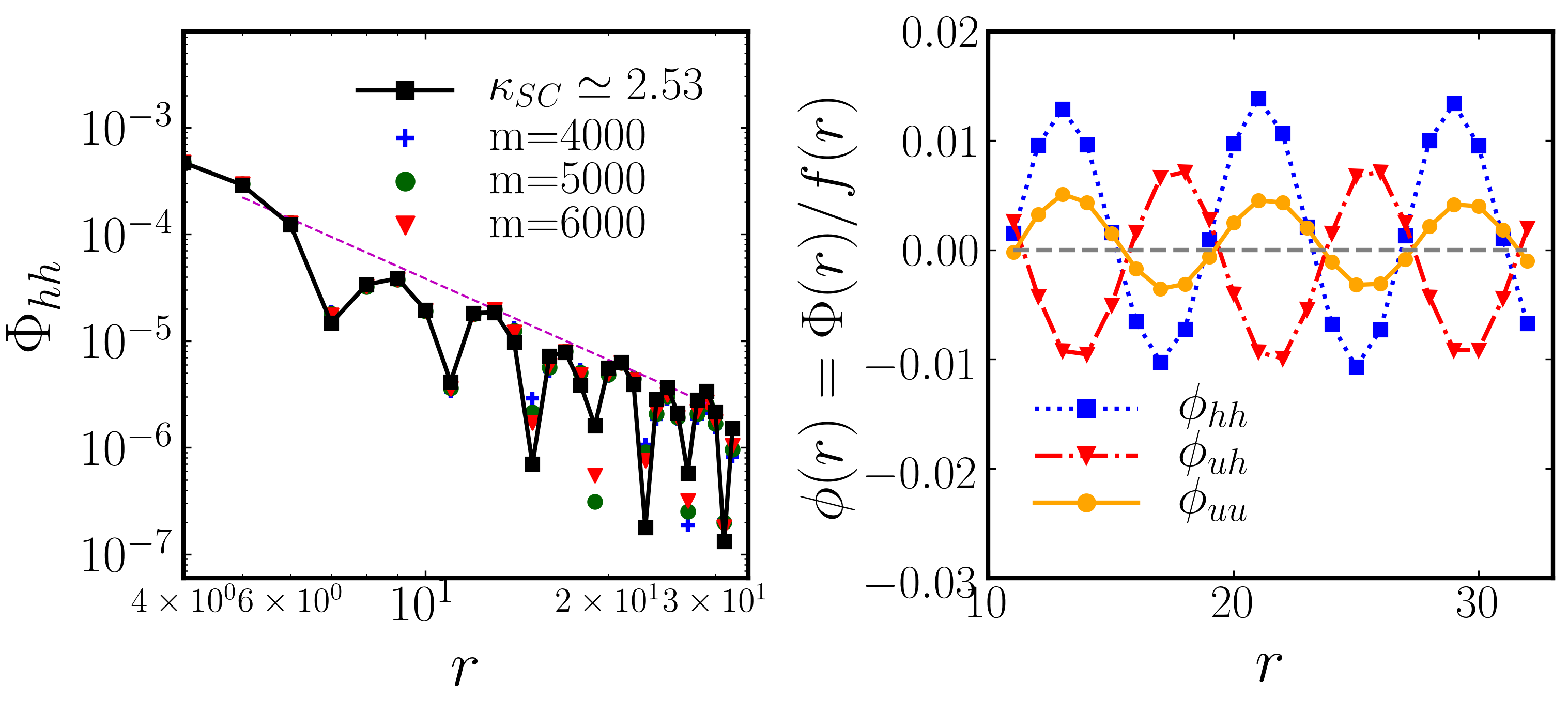}
\caption{Correlations for $t_{dd}=t_{pp}^{\prime}=0$. Left: Singlet Pairing correlations on h bonds in log scale, the black curve with square symbols represents the result of extrapolation to $m\rightarrow \infty$; Right: the normalized pairing correlations on $h-h, u-u,$ and $u-h$ bonds. $f(r)$ is the envelope function. }
\label{fig:corr_LL_pair}
\end{figure}

\begin{figure}
\centering
\includegraphics[width=.45\textwidth]{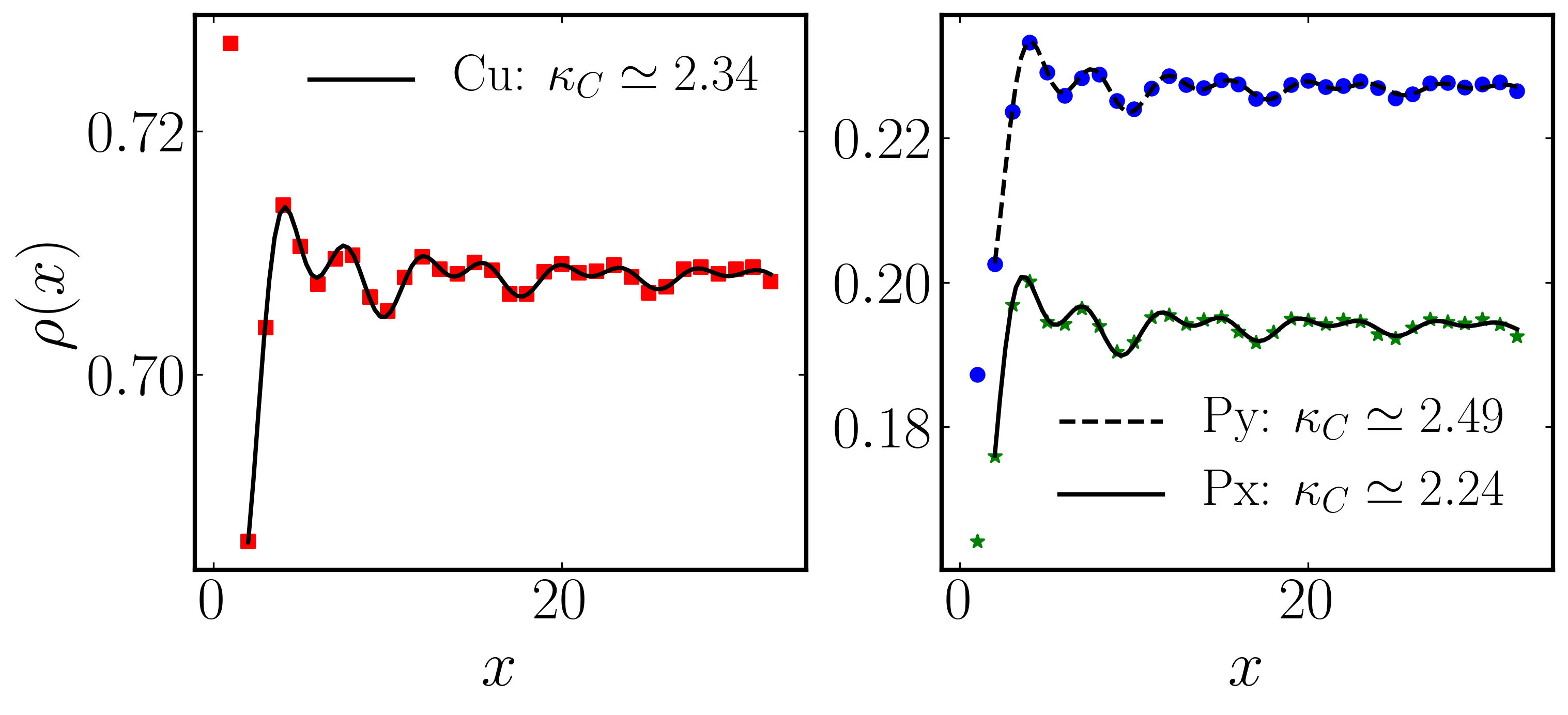}
\caption{Local density profiles for $t_{dd}=t_{pp}^{\prime}=0$. Left: density distribution on copper orbital; Right: density distribution on oxygen orbital.  $\kappa_C$ is the scaling exponent obtained by curve fitting.}
\label{fig:corr_LL_local}
\end{figure}

We choose one representative data point in the LL phase and present the ground state properties. For the case $t_{dd}=t_{pp}^{\prime}=0$, we show the pairing correlations in Fig. \ref{fig:corr_LL_pair}. The pairing order has a spatial oscillation with a vanishing average, and exhibits a $d
$-wave symmetry. Fig. \ref{fig:corr_LL_local} shows the local density profile and the decaying exponent obtained by curve fitting with Eq. \ref{n_profile}.

\clearpage

\bibliography{main.bib}

\begin{thebibliography}{50}%
\makeatletter
\providecommand \@ifxundefined [1]{%
 \@ifx{#1\undefined}
}%
\providecommand \@ifnum [1]{%
 \ifnum #1\expandafter \@firstoftwo
 \else \expandafter \@secondoftwo
 \fi
}%
\providecommand \@ifx [1]{%
 \ifx #1\expandafter \@firstoftwo
 \else \expandafter \@secondoftwo
 \fi
}%
\providecommand \natexlab [1]{#1}%
\providecommand \enquote  [1]{``#1''}%
\providecommand \bibnamefont  [1]{#1}%
\providecommand \bibfnamefont [1]{#1}%
\providecommand \citenamefont [1]{#1}%
\providecommand \href@noop [0]{\@secondoftwo}%
\providecommand \href [0]{\begingroup \@sanitize@url \@href}%
\providecommand \@href[1]{\@@startlink{#1}\@@href}%
\providecommand \@@href[1]{\endgroup#1\@@endlink}%
\providecommand \@sanitize@url [0]{\catcode `\\12\catcode `\$12\catcode
  `\&12\catcode `\#12\catcode `\^12\catcode `\_12\catcode `\%12\relax}%
\providecommand \@@startlink[1]{}%
\providecommand \@@endlink[0]{}%
\providecommand \url  [0]{\begingroup\@sanitize@url \@url }%
\providecommand \@url [1]{\endgroup\@href {#1}{\urlprefix }}%
\providecommand \urlprefix  [0]{URL }%
\providecommand \Eprint [0]{\href }%
\providecommand \doibase [0]{https://doi.org/}%
\providecommand \selectlanguage [0]{\@gobble}%
\providecommand \bibinfo  [0]{\@secondoftwo}%
\providecommand \bibfield  [0]{\@secondoftwo}%
\providecommand \translation [1]{[#1]}%
\providecommand \BibitemOpen [0]{}%
\providecommand \bibitemStop [0]{}%
\providecommand \bibitemNoStop [0]{.\EOS\space}%
\providecommand \EOS [0]{\spacefactor3000\relax}%
\providecommand \BibitemShut  [1]{\csname bibitem#1\endcsname}%
\let\auto@bib@innerbib\@empty
\bibitem [{\citenamefont {Emery}(1987)}]{Emery1987}%
  \BibitemOpen
  \bibfield  {author} {\bibinfo {author} {\bibfnamefont {V.~J.}\ \bibnamefont
  {Emery}},\ }\bibfield  {title} {\bibinfo {title} {Theory of
  high-${\mathrm{t}}_{\mathrm{c}}$ superconductivity in oxides},\ }\href
  {https://doi.org/10.1103/PhysRevLett.58.2794} {\bibfield  {journal} {\bibinfo
   {journal} {Phys. Rev. Lett.}\ }\textbf {\bibinfo {volume} {58}},\ \bibinfo
  {pages} {2794} (\bibinfo {year} {1987})}\BibitemShut {NoStop}%
\bibitem [{\citenamefont {Emery}\ and\ \citenamefont
  {Reiter}(1988)}]{Emery1988}%
  \BibitemOpen
  \bibfield  {author} {\bibinfo {author} {\bibfnamefont {V.~J.}\ \bibnamefont
  {Emery}}\ and\ \bibinfo {author} {\bibfnamefont {G.}~\bibnamefont {Reiter}},\
  }\bibfield  {title} {\bibinfo {title} {Mechanism for high-temperature
  superconductivity},\ }\href {https://doi.org/10.1103/PhysRevB.38.4547}
  {\bibfield  {journal} {\bibinfo  {journal} {Phys. Rev. B}\ }\textbf {\bibinfo
  {volume} {38}},\ \bibinfo {pages} {4547} (\bibinfo {year}
  {1988})}\BibitemShut {NoStop}%
\bibitem [{\citenamefont {Jiang}(2023)}]{Jiang2023}%
  \BibitemOpen
  \bibfield  {author} {\bibinfo {author} {\bibfnamefont {H.-C.}\ \bibnamefont
  {Jiang}},\ }\bibfield  {title} {\bibinfo {title} {Pair density wave in the
  doped three-band hubbard model on two-leg square cylinders},\ }\href
  {https://doi.org/10.1103/PhysRevB.107.214504} {\bibfield  {journal} {\bibinfo
   {journal} {Phys. Rev. B}\ }\textbf {\bibinfo {volume} {107}},\ \bibinfo
  {pages} {214504} (\bibinfo {year} {2023})}\BibitemShut {NoStop}%
\bibitem [{\citenamefont {Jiang}\ and\ \citenamefont
  {Devereaux}(2023)}]{jiang2023_2}%
  \BibitemOpen
  \bibfield  {author} {\bibinfo {author} {\bibfnamefont {H.-C.}\ \bibnamefont
  {Jiang}}\ and\ \bibinfo {author} {\bibfnamefont {T.~P.}\ \bibnamefont
  {Devereaux}},\ }\href@noop {} {\bibinfo {title} {Pair density wave and
  superconductivity in a kinetically frustrated doped emery model on a square
  lattice}} (\bibinfo {year} {2023}),\ \Eprint
  {https://arxiv.org/abs/2309.11786} {arXiv:2309.11786 [cond-mat.str-el]}
  \BibitemShut {NoStop}%
\bibitem [{\citenamefont {Berg}\ \emph {et~al.}(2010)\citenamefont {Berg},
  \citenamefont {Fradkin},\ and\ \citenamefont {Kivelson}}]{Berg2010}%
  \BibitemOpen
  \bibfield  {author} {\bibinfo {author} {\bibfnamefont {E.}~\bibnamefont
  {Berg}}, \bibinfo {author} {\bibfnamefont {E.}~\bibnamefont {Fradkin}},\ and\
  \bibinfo {author} {\bibfnamefont {S.~A.}\ \bibnamefont {Kivelson}},\
  }\bibfield  {title} {\bibinfo {title} {Pair-density-wave correlations in the
  kondo-heisenberg model},\ }\href
  {https://doi.org/10.1103/PhysRevLett.105.146403} {\bibfield  {journal}
  {\bibinfo  {journal} {Phys. Rev. Lett.}\ }\textbf {\bibinfo {volume} {105}},\
  \bibinfo {pages} {146403} (\bibinfo {year} {2010})}\BibitemShut {NoStop}%
\bibitem [{\citenamefont {Almeida}\ \emph {et~al.}(2010)\citenamefont
  {Almeida}, \citenamefont {Roux},\ and\ \citenamefont
  {Poilblanc}}]{Almeida2010}%
  \BibitemOpen
  \bibfield  {author} {\bibinfo {author} {\bibfnamefont {J.}~\bibnamefont
  {Almeida}}, \bibinfo {author} {\bibfnamefont {G.}~\bibnamefont {Roux}},\ and\
  \bibinfo {author} {\bibfnamefont {D.}~\bibnamefont {Poilblanc}},\ }\bibfield
  {title} {\bibinfo {title} {Pair density waves in coupled doped two-leg
  ladders},\ }\href {https://doi.org/10.1103/PhysRevB.82.041102} {\bibfield
  {journal} {\bibinfo  {journal} {Phys. Rev. B}\ }\textbf {\bibinfo {volume}
  {82}},\ \bibinfo {pages} {041102} (\bibinfo {year} {2010})}\BibitemShut
  {NoStop}%
\bibitem [{\citenamefont {Jaefari}\ and\ \citenamefont
  {Fradkin}(2012)}]{Jaefari2012}%
  \BibitemOpen
  \bibfield  {author} {\bibinfo {author} {\bibfnamefont {A.}~\bibnamefont
  {Jaefari}}\ and\ \bibinfo {author} {\bibfnamefont {E.}~\bibnamefont
  {Fradkin}},\ }\bibfield  {title} {\bibinfo {title} {Pair-density-wave
  superconducting order in two-leg ladders},\ }\href
  {https://doi.org/10.1103/PhysRevB.85.035104} {\bibfield  {journal} {\bibinfo
  {journal} {Phys. Rev. B}\ }\textbf {\bibinfo {volume} {85}},\ \bibinfo
  {pages} {035104} (\bibinfo {year} {2012})}\BibitemShut {NoStop}%
\bibitem [{\citenamefont {Zegrodnik}\ and\ \citenamefont
  {Spa\l{}ek}(2018)}]{Zegrodnik2018}%
  \BibitemOpen
  \bibfield  {author} {\bibinfo {author} {\bibfnamefont {M.}~\bibnamefont
  {Zegrodnik}}\ and\ \bibinfo {author} {\bibfnamefont {J.}~\bibnamefont
  {Spa\l{}ek}},\ }\bibfield  {title} {\bibinfo {title} {Incorporation of
  charge- and pair-density-wave states into the one-band model of $d$-wave
  superconductivity},\ }\href {https://doi.org/10.1103/PhysRevB.98.155144}
  {\bibfield  {journal} {\bibinfo  {journal} {Phys. Rev. B}\ }\textbf {\bibinfo
  {volume} {98}},\ \bibinfo {pages} {155144} (\bibinfo {year}
  {2018})}\BibitemShut {NoStop}%
\bibitem [{\citenamefont {Xu}\ \emph {et~al.}(2019)\citenamefont {Xu},
  \citenamefont {Law},\ and\ \citenamefont {Lee}}]{Xu2019}%
  \BibitemOpen
  \bibfield  {author} {\bibinfo {author} {\bibfnamefont {X.~Y.}\ \bibnamefont
  {Xu}}, \bibinfo {author} {\bibfnamefont {K.~T.}\ \bibnamefont {Law}},\ and\
  \bibinfo {author} {\bibfnamefont {P.~A.}\ \bibnamefont {Lee}},\ }\bibfield
  {title} {\bibinfo {title} {Pair density wave in the doped
  $t\text{\ensuremath{-}}j$ model with ring exchange on a triangular lattice},\
  }\href {https://doi.org/10.1103/PhysRevLett.122.167001} {\bibfield  {journal}
  {\bibinfo  {journal} {Phys. Rev. Lett.}\ }\textbf {\bibinfo {volume} {122}},\
  \bibinfo {pages} {167001} (\bibinfo {year} {2019})}\BibitemShut {NoStop}%
\bibitem [{\citenamefont {May-Mann}\ \emph {et~al.}(2020)\citenamefont
  {May-Mann}, \citenamefont {Levy}, \citenamefont {Soto-Garrido}, \citenamefont
  {Cho}, \citenamefont {Clark},\ and\ \citenamefont {Fradkin}}]{May2020}%
  \BibitemOpen
  \bibfield  {author} {\bibinfo {author} {\bibfnamefont {J.}~\bibnamefont
  {May-Mann}}, \bibinfo {author} {\bibfnamefont {R.}~\bibnamefont {Levy}},
  \bibinfo {author} {\bibfnamefont {R.}~\bibnamefont {Soto-Garrido}}, \bibinfo
  {author} {\bibfnamefont {G.~Y.}\ \bibnamefont {Cho}}, \bibinfo {author}
  {\bibfnamefont {B.~K.}\ \bibnamefont {Clark}},\ and\ \bibinfo {author}
  {\bibfnamefont {E.}~\bibnamefont {Fradkin}},\ }\bibfield  {title} {\bibinfo
  {title} {Topology and the one-dimensional kondo-heisenberg model},\ }\href
  {https://doi.org/10.1103/PhysRevB.101.165133} {\bibfield  {journal} {\bibinfo
   {journal} {Phys. Rev. B}\ }\textbf {\bibinfo {volume} {101}},\ \bibinfo
  {pages} {165133} (\bibinfo {year} {2020})}\BibitemShut {NoStop}%
\bibitem [{\citenamefont {Zhang}\ and\ \citenamefont
  {Vishwanath}(2022)}]{Zhang2022}%
  \BibitemOpen
  \bibfield  {author} {\bibinfo {author} {\bibfnamefont {Y.-H.}\ \bibnamefont
  {Zhang}}\ and\ \bibinfo {author} {\bibfnamefont {A.}~\bibnamefont
  {Vishwanath}},\ }\bibfield  {title} {\bibinfo {title} {Pair-density-wave
  superconductor from doping haldane chain and rung-singlet ladder},\ }\href
  {https://doi.org/10.1103/PhysRevB.106.045103} {\bibfield  {journal} {\bibinfo
   {journal} {Phys. Rev. B}\ }\textbf {\bibinfo {volume} {106}},\ \bibinfo
  {pages} {045103} (\bibinfo {year} {2022})}\BibitemShut {NoStop}%
\bibitem [{\citenamefont {Arovas}\ \emph {et~al.}(2022)\citenamefont {Arovas},
  \citenamefont {Berg}, \citenamefont {Kivelson},\ and\ \citenamefont
  {Raghu}}]{Arovas2022}%
  \BibitemOpen
  \bibfield  {author} {\bibinfo {author} {\bibfnamefont {D.~P.}\ \bibnamefont
  {Arovas}}, \bibinfo {author} {\bibfnamefont {E.}~\bibnamefont {Berg}},
  \bibinfo {author} {\bibfnamefont {S.~A.}\ \bibnamefont {Kivelson}},\ and\
  \bibinfo {author} {\bibfnamefont {S.}~\bibnamefont {Raghu}},\ }\bibfield
  {title} {\bibinfo {title} {The hubbard model},\ }\href
  {https://doi.org/10.1146/annurev-conmatphys-031620-102024} {\bibfield
  {journal} {\bibinfo  {journal} {Annual Review of Condensed Matter Physics}\
  }\textbf {\bibinfo {volume} {13}},\ \bibinfo {pages} {239} (\bibinfo {year}
  {2022})},\ \Eprint
  {https://arxiv.org/abs/https://doi.org/10.1146/annurev-conmatphys-031620-102024}
  {https://doi.org/10.1146/annurev-conmatphys-031620-102024} \BibitemShut
  {NoStop}%
\bibitem [{\citenamefont {Qin}\ \emph {et~al.}(2022)\citenamefont {Qin},
  \citenamefont {Sch\"{a}fer}, \citenamefont {Andergassen}, \citenamefont
  {Corboz},\ and\ \citenamefont {Gull}}]{Qin2022}%
  \BibitemOpen
  \bibfield  {author} {\bibinfo {author} {\bibfnamefont {M.}~\bibnamefont
  {Qin}}, \bibinfo {author} {\bibfnamefont {T.}~\bibnamefont {Sch\"{a}fer}},
  \bibinfo {author} {\bibfnamefont {S.}~\bibnamefont {Andergassen}}, \bibinfo
  {author} {\bibfnamefont {P.}~\bibnamefont {Corboz}},\ and\ \bibinfo {author}
  {\bibfnamefont {E.}~\bibnamefont {Gull}},\ }\bibfield  {title} {\bibinfo
  {title} {The hubbard model: A computational perspective},\ }\href
  {https://doi.org/10.1146/annurev-conmatphys-090921-033948} {\bibfield
  {journal} {\bibinfo  {journal} {Annual Review of Condensed Matter Physics}\
  }\textbf {\bibinfo {volume} {13}},\ \bibinfo {pages} {275} (\bibinfo {year}
  {2022})},\ \Eprint
  {https://arxiv.org/abs/https://doi.org/10.1146/annurev-conmatphys-090921-033948}
  {https://doi.org/10.1146/annurev-conmatphys-090921-033948} \BibitemShut
  {NoStop}%
\bibitem [{\citenamefont {Zhang}\ and\ \citenamefont {Rice}(1988)}]{zhangrice}%
  \BibitemOpen
  \bibfield  {author} {\bibinfo {author} {\bibfnamefont {F.~C.}\ \bibnamefont
  {Zhang}}\ and\ \bibinfo {author} {\bibfnamefont {T.~M.}\ \bibnamefont
  {Rice}},\ }\bibfield  {title} {\bibinfo {title} {Effective hamiltonian for
  the superconducting cu oxides},\ }\href
  {https://doi.org/10.1103/PhysRevB.37.3759} {\bibfield  {journal} {\bibinfo
  {journal} {Phys. Rev. B}\ }\textbf {\bibinfo {volume} {37}},\ \bibinfo
  {pages} {3759} (\bibinfo {year} {1988})}\BibitemShut {NoStop}%
\bibitem [{\citenamefont {Jiang}\ \emph
  {et~al.}(2023{\natexlab{a}})\citenamefont {Jiang}, \citenamefont
  {Scalapino},\ and\ \citenamefont {White}}]{jiang2023singleband}%
  \BibitemOpen
  \bibfield  {author} {\bibinfo {author} {\bibfnamefont {S.}~\bibnamefont
  {Jiang}}, \bibinfo {author} {\bibfnamefont {D.~J.}\ \bibnamefont
  {Scalapino}},\ and\ \bibinfo {author} {\bibfnamefont {S.~R.}\ \bibnamefont
  {White}},\ }\href@noop {} {\bibinfo {title} {A single-band model with
  enhanced pairing from dmrg-based downfolding of the three-band hubbard
  model}} (\bibinfo {year} {2023}{\natexlab{a}}),\ \Eprint
  {https://arxiv.org/abs/2303.00756} {arXiv:2303.00756 [cond-mat.str-el]}
  \BibitemShut {NoStop}%
\bibitem [{\citenamefont {Eskes}\ \emph {et~al.}(1989)\citenamefont {Eskes},
  \citenamefont {Sawatzky},\ and\ \citenamefont {Feiner}}]{ESKES1989424}%
  \BibitemOpen
  \bibfield  {author} {\bibinfo {author} {\bibfnamefont {H.}~\bibnamefont
  {Eskes}}, \bibinfo {author} {\bibfnamefont {G.}~\bibnamefont {Sawatzky}},\
  and\ \bibinfo {author} {\bibfnamefont {L.}~\bibnamefont {Feiner}},\
  }\bibfield  {title} {\bibinfo {title} {Effective transfer for singlets formed
  by hole doping in the high-tc superconductors},\ }\href
  {https://doi.org/https://doi.org/10.1016/0921-4534(89)90415-2} {\bibfield
  {journal} {\bibinfo  {journal} {Physica C: Superconductivity}\ }\textbf
  {\bibinfo {volume} {160}},\ \bibinfo {pages} {424} (\bibinfo {year}
  {1989})}\BibitemShut {NoStop}%
\bibitem [{\citenamefont {Eskes}\ and\ \citenamefont
  {Sawatzky}(1991{\natexlab{a}})}]{Eskes1991}%
  \BibitemOpen
  \bibfield  {author} {\bibinfo {author} {\bibfnamefont {H.}~\bibnamefont
  {Eskes}}\ and\ \bibinfo {author} {\bibfnamefont {G.~A.}\ \bibnamefont
  {Sawatzky}},\ }\bibfield  {title} {\bibinfo {title} {Single-, triple-, or
  multiple-band hubbard models},\ }\href
  {https://doi.org/10.1103/PhysRevB.44.9656} {\bibfield  {journal} {\bibinfo
  {journal} {Phys. Rev. B}\ }\textbf {\bibinfo {volume} {44}},\ \bibinfo
  {pages} {9656} (\bibinfo {year} {1991}{\natexlab{a}})}\BibitemShut {NoStop}%
\bibitem [{\citenamefont {Eskes}\ and\ \citenamefont
  {Sawatzky}(1991{\natexlab{b}})}]{Eskes1991_2}%
  \BibitemOpen
  \bibfield  {author} {\bibinfo {author} {\bibfnamefont {H.}~\bibnamefont
  {Eskes}}\ and\ \bibinfo {author} {\bibfnamefont {G.~A.}\ \bibnamefont
  {Sawatzky}},\ }\bibfield  {title} {\bibinfo {title} {Doping dependence of
  high-energy spectral weights for the high-${\mathit{t}}_{\mathit{c}}$
  cuprates},\ }\href {https://doi.org/10.1103/PhysRevB.43.119} {\bibfield
  {journal} {\bibinfo  {journal} {Phys. Rev. B}\ }\textbf {\bibinfo {volume}
  {43}},\ \bibinfo {pages} {119} (\bibinfo {year}
  {1991}{\natexlab{b}})}\BibitemShut {NoStop}%
\bibitem [{\citenamefont {Feiner}\ \emph {et~al.}(1996)\citenamefont {Feiner},
  \citenamefont {Jefferson},\ and\ \citenamefont {Raimondi}}]{Raimondi1996_1}%
  \BibitemOpen
  \bibfield  {author} {\bibinfo {author} {\bibfnamefont {L.~F.}\ \bibnamefont
  {Feiner}}, \bibinfo {author} {\bibfnamefont {J.~H.}\ \bibnamefont
  {Jefferson}},\ and\ \bibinfo {author} {\bibfnamefont {R.}~\bibnamefont
  {Raimondi}},\ }\bibfield  {title} {\bibinfo {title} {Effective single-band
  models for the high-${\mathit{t}}_{\mathit{c}}$ cuprates. i. coulomb
  interactions},\ }\href {https://doi.org/10.1103/PhysRevB.53.8751} {\bibfield
  {journal} {\bibinfo  {journal} {Phys. Rev. B}\ }\textbf {\bibinfo {volume}
  {53}},\ \bibinfo {pages} {8751} (\bibinfo {year} {1996})}\BibitemShut
  {NoStop}%
\bibitem [{\citenamefont {Raimondi}\ \emph {et~al.}(1996)\citenamefont
  {Raimondi}, \citenamefont {Jefferson},\ and\ \citenamefont
  {Feiner}}]{Raimondi1996_2}%
  \BibitemOpen
  \bibfield  {author} {\bibinfo {author} {\bibfnamefont {R.}~\bibnamefont
  {Raimondi}}, \bibinfo {author} {\bibfnamefont {J.~H.}\ \bibnamefont
  {Jefferson}},\ and\ \bibinfo {author} {\bibfnamefont {L.~F.}\ \bibnamefont
  {Feiner}},\ }\bibfield  {title} {\bibinfo {title} {Effective single-band
  models for the high-${\mathit{t}}_{\mathit{c}}$ cuprates. ii. role of apical
  oxygen},\ }\href {https://doi.org/10.1103/PhysRevB.53.8774} {\bibfield
  {journal} {\bibinfo  {journal} {Phys. Rev. B}\ }\textbf {\bibinfo {volume}
  {53}},\ \bibinfo {pages} {8774} (\bibinfo {year} {1996})}\BibitemShut
  {NoStop}%
\bibitem [{\citenamefont {Belinicher}\ and\ \citenamefont
  {Chernyshev}(1993)}]{Belinicher1993}%
  \BibitemOpen
  \bibfield  {author} {\bibinfo {author} {\bibfnamefont {V.~I.}\ \bibnamefont
  {Belinicher}}\ and\ \bibinfo {author} {\bibfnamefont {A.~L.}\ \bibnamefont
  {Chernyshev}},\ }\bibfield  {title} {\bibinfo {title} {Reduction of a
  three-band model for copper oxides to a single-band generalized t-j model},\
  }\href {https://doi.org/10.1103/PhysRevB.47.390} {\bibfield  {journal}
  {\bibinfo  {journal} {Phys. Rev. B}\ }\textbf {\bibinfo {volume} {47}},\
  \bibinfo {pages} {390} (\bibinfo {year} {1993})}\BibitemShut {NoStop}%
\bibitem [{\citenamefont {Huang}\ \emph {et~al.}(2017)\citenamefont {Huang},
  \citenamefont {Mendl}, \citenamefont {Liu}, \citenamefont {Johnston},
  \citenamefont {Jiang}, \citenamefont {Moritz},\ and\ \citenamefont
  {Devereaux}}]{Huang2017}%
  \BibitemOpen
  \bibfield  {author} {\bibinfo {author} {\bibfnamefont {E.~W.}\ \bibnamefont
  {Huang}}, \bibinfo {author} {\bibfnamefont {C.~B.}\ \bibnamefont {Mendl}},
  \bibinfo {author} {\bibfnamefont {S.}~\bibnamefont {Liu}}, \bibinfo {author}
  {\bibfnamefont {S.}~\bibnamefont {Johnston}}, \bibinfo {author}
  {\bibfnamefont {H.-C.}\ \bibnamefont {Jiang}}, \bibinfo {author}
  {\bibfnamefont {B.}~\bibnamefont {Moritz}},\ and\ \bibinfo {author}
  {\bibfnamefont {T.~P.}\ \bibnamefont {Devereaux}},\ }\bibfield  {title}
  {\bibinfo {title} {Numerical evidence of fluctuating stripes in the normal
  state of high-tc cuprate superconductors},\ }\href
  {https://www.science.org/doi/abs/10.1126/science.aak9546} {\bibfield
  {journal} {\bibinfo  {journal} {Science}\ }\textbf {\bibinfo {volume}
  {358}},\ \bibinfo {pages} {1161} (\bibinfo {year} {2017})}\BibitemShut
  {NoStop}%
\bibitem [{\citenamefont {Huang}\ \emph {et~al.}(2018)\citenamefont {Huang},
  \citenamefont {Mendl}, \citenamefont {Jiang}, \citenamefont {Moritz},\ and\
  \citenamefont {Devereaux}}]{Huang2018}%
  \BibitemOpen
  \bibfield  {author} {\bibinfo {author} {\bibfnamefont {E.~W.}\ \bibnamefont
  {Huang}}, \bibinfo {author} {\bibfnamefont {C.~B.}\ \bibnamefont {Mendl}},
  \bibinfo {author} {\bibfnamefont {H.-C.}\ \bibnamefont {Jiang}}, \bibinfo
  {author} {\bibfnamefont {B.}~\bibnamefont {Moritz}},\ and\ \bibinfo {author}
  {\bibfnamefont {T.~P.}\ \bibnamefont {Devereaux}},\ }\bibfield  {title}
  {\bibinfo {title} {Stripe order from the perspective of the hubbard model},\
  }\href {https://www.nature.com/articles/s41535-018-0097-0#citeas} {\bibfield
  {journal} {\bibinfo  {journal} {Nature News}\ } (\bibinfo {year}
  {2018})}\BibitemShut {NoStop}%
\bibitem [{\citenamefont {Hirsch}(1985)}]{Hirsch1985}%
  \BibitemOpen
  \bibfield  {author} {\bibinfo {author} {\bibfnamefont {J.~E.}\ \bibnamefont
  {Hirsch}},\ }\bibfield  {title} {\bibinfo {title} {Two-dimensional hubbard
  model: Numerical simulation study},\ }\href
  {https://doi.org/10.1103/PhysRevB.31.4403} {\bibfield  {journal} {\bibinfo
  {journal} {Phys. Rev. B}\ }\textbf {\bibinfo {volume} {31}},\ \bibinfo
  {pages} {4403} (\bibinfo {year} {1985})}\BibitemShut {NoStop}%
\bibitem [{\citenamefont {Yokoyama}\ \emph {et~al.}(2006)\citenamefont
  {Yokoyama}, \citenamefont {Ogata},\ and\ \citenamefont
  {Tanaka}}]{Yokoma2006}%
  \BibitemOpen
  \bibfield  {author} {\bibinfo {author} {\bibfnamefont {H.}~\bibnamefont
  {Yokoyama}}, \bibinfo {author} {\bibfnamefont {M.}~\bibnamefont {Ogata}},\
  and\ \bibinfo {author} {\bibfnamefont {Y.}~\bibnamefont {Tanaka}},\
  }\bibfield  {title} {\bibinfo {title} {Mott transitions and d-wave
  superconductivity in half-filled-band hubbard model on square lattice with
  geometric frustration},\ }\href {https://doi.org/10.1143/JPSJ.75.114706}
  {\bibfield  {journal} {\bibinfo  {journal} {Journal of the Physical Society
  of Japan}\ }\textbf {\bibinfo {volume} {75}},\ \bibinfo {pages} {114706}
  (\bibinfo {year} {2006})},\ \Eprint
  {https://arxiv.org/abs/https://doi.org/10.1143/JPSJ.75.114706}
  {https://doi.org/10.1143/JPSJ.75.114706} \BibitemShut {NoStop}%
\bibitem [{\citenamefont {Dopf}\ \emph {et~al.}(1990)\citenamefont {Dopf},
  \citenamefont {Muramatsu},\ and\ \citenamefont {Hanke}}]{Dopf1990}%
  \BibitemOpen
  \bibfield  {author} {\bibinfo {author} {\bibfnamefont {G.}~\bibnamefont
  {Dopf}}, \bibinfo {author} {\bibfnamefont {A.}~\bibnamefont {Muramatsu}},\
  and\ \bibinfo {author} {\bibfnamefont {W.}~\bibnamefont {Hanke}},\ }\bibfield
   {title} {\bibinfo {title} {Three-band hubbard model: A monte carlo study},\
  }\href {https://doi.org/10.1103/PhysRevB.41.9264} {\bibfield  {journal}
  {\bibinfo  {journal} {Phys. Rev. B}\ }\textbf {\bibinfo {volume} {41}},\
  \bibinfo {pages} {9264} (\bibinfo {year} {1990})}\BibitemShut {NoStop}%
\bibitem [{\citenamefont {Yanagisawa}\ \emph {et~al.}(2001)\citenamefont
  {Yanagisawa}, \citenamefont {Koike},\ and\ \citenamefont
  {Yamaji}}]{Yanagisawa2001}%
  \BibitemOpen
  \bibfield  {author} {\bibinfo {author} {\bibfnamefont {T.}~\bibnamefont
  {Yanagisawa}}, \bibinfo {author} {\bibfnamefont {S.}~\bibnamefont {Koike}},\
  and\ \bibinfo {author} {\bibfnamefont {K.}~\bibnamefont {Yamaji}},\
  }\bibfield  {title} {\bibinfo {title} {Ground state of the three-band hubbard
  model},\ }\href {https://doi.org/10.1103/PhysRevB.64.184509} {\bibfield
  {journal} {\bibinfo  {journal} {Phys. Rev. B}\ }\textbf {\bibinfo {volume}
  {64}},\ \bibinfo {pages} {184509} (\bibinfo {year} {2001})}\BibitemShut
  {NoStop}%
\bibitem [{\citenamefont {Shen}\ \emph {et~al.}(2023)\citenamefont {Shen},
  \citenamefont {Zhang},\ and\ \citenamefont {Qin}}]{ShenYang2023}%
  \BibitemOpen
  \bibfield  {author} {\bibinfo {author} {\bibfnamefont {Y.}~\bibnamefont
  {Shen}}, \bibinfo {author} {\bibfnamefont {G.-M.}\ \bibnamefont {Zhang}},\
  and\ \bibinfo {author} {\bibfnamefont {M.}~\bibnamefont {Qin}},\ }\bibfield
  {title} {\bibinfo {title} {Reexamining doped two-legged hubbard ladders},\
  }\href {https://doi.org/10.1103/PhysRevB.108.165113} {\bibfield  {journal}
  {\bibinfo  {journal} {Phys. Rev. B}\ }\textbf {\bibinfo {volume} {108}},\
  \bibinfo {pages} {165113} (\bibinfo {year} {2023})}\BibitemShut {NoStop}%
\bibitem [{\citenamefont {Jiang}\ \emph {et~al.}(2018)\citenamefont {Jiang},
  \citenamefont {Weng},\ and\ \citenamefont {Kivelson}}]{Hongchen2018}%
  \BibitemOpen
  \bibfield  {author} {\bibinfo {author} {\bibfnamefont {H.-C.}\ \bibnamefont
  {Jiang}}, \bibinfo {author} {\bibfnamefont {Z.-Y.}\ \bibnamefont {Weng}},\
  and\ \bibinfo {author} {\bibfnamefont {S.~A.}\ \bibnamefont {Kivelson}},\
  }\bibfield  {title} {\bibinfo {title} {Superconductivity in the doped
  $\mathit{t}\ensuremath{-}\mathit{J}$ model: Results for four-leg cylinders},\
  }\href {https://doi.org/10.1103/PhysRevB.98.140505} {\bibfield  {journal}
  {\bibinfo  {journal} {Phys. Rev. B}\ }\textbf {\bibinfo {volume} {98}},\
  \bibinfo {pages} {140505} (\bibinfo {year} {2018})}\BibitemShut {NoStop}%
\bibitem [{\citenamefont {Song}\ \emph {et~al.}(2021)\citenamefont {Song},
  \citenamefont {Mazumdar},\ and\ \citenamefont {Clay}}]{Song2021}%
  \BibitemOpen
  \bibfield  {author} {\bibinfo {author} {\bibfnamefont {J.-P.}\ \bibnamefont
  {Song}}, \bibinfo {author} {\bibfnamefont {S.}~\bibnamefont {Mazumdar}},\
  and\ \bibinfo {author} {\bibfnamefont {R.~T.}\ \bibnamefont {Clay}},\
  }\bibfield  {title} {\bibinfo {title} {Absence of luther-emery
  superconducting phase in the three-band model for cuprate ladders},\ }\href
  {https://doi.org/10.1103/PhysRevB.104.104504} {\bibfield  {journal} {\bibinfo
   {journal} {Phys. Rev. B}\ }\textbf {\bibinfo {volume} {104}},\ \bibinfo
  {pages} {104504} (\bibinfo {year} {2021})}\BibitemShut {NoStop}%
\bibitem [{\citenamefont {Aichhorn}\ \emph {et~al.}(2006)\citenamefont
  {Aichhorn}, \citenamefont {Arrigoni}, \citenamefont {Potthoff},\ and\
  \citenamefont {Hanke}}]{Aichhorn2006}%
  \BibitemOpen
  \bibfield  {author} {\bibinfo {author} {\bibfnamefont {M.}~\bibnamefont
  {Aichhorn}}, \bibinfo {author} {\bibfnamefont {E.}~\bibnamefont {Arrigoni}},
  \bibinfo {author} {\bibfnamefont {M.}~\bibnamefont {Potthoff}},\ and\
  \bibinfo {author} {\bibfnamefont {W.}~\bibnamefont {Hanke}},\ }\bibfield
  {title} {\bibinfo {title} {Antiferromagnetic to superconducting phase
  transition in the hole- and electron-doped hubbard model at zero
  temperature},\ }\href {https://doi.org/10.1103/PhysRevB.74.024508} {\bibfield
   {journal} {\bibinfo  {journal} {Phys. Rev. B}\ }\textbf {\bibinfo {volume}
  {74}},\ \bibinfo {pages} {024508} (\bibinfo {year} {2006})}\BibitemShut
  {NoStop}%
\bibitem [{\citenamefont {Ponsioen}\ \emph {et~al.}(2019)\citenamefont
  {Ponsioen}, \citenamefont {Chung},\ and\ \citenamefont
  {Corboz}}]{Ponsioen2019}%
  \BibitemOpen
  \bibfield  {author} {\bibinfo {author} {\bibfnamefont {B.}~\bibnamefont
  {Ponsioen}}, \bibinfo {author} {\bibfnamefont {S.~S.}\ \bibnamefont
  {Chung}},\ and\ \bibinfo {author} {\bibfnamefont {P.}~\bibnamefont
  {Corboz}},\ }\bibfield  {title} {\bibinfo {title} {Period 4 stripe in the
  extended two-dimensional hubbard model},\ }\href
  {https://doi.org/10.1103/PhysRevB.100.195141} {\bibfield  {journal} {\bibinfo
   {journal} {Phys. Rev. B}\ }\textbf {\bibinfo {volume} {100}},\ \bibinfo
  {pages} {195141} (\bibinfo {year} {2019})}\BibitemShut {NoStop}%
\bibitem [{\citenamefont {Jiang}\ and\ \citenamefont
  {Devereaux}(2019)}]{Jiang2019Hub}%
  \BibitemOpen
  \bibfield  {author} {\bibinfo {author} {\bibfnamefont {H.-C.}\ \bibnamefont
  {Jiang}}\ and\ \bibinfo {author} {\bibfnamefont {T.~P.}\ \bibnamefont
  {Devereaux}},\ }\bibfield  {title} {\bibinfo {title} {Superconductivity in
  the doped hubbard model and its interplay with next-nearest hopping t'},\
  }\href {https://doi.org/10.1126/science.aal5304} {\bibfield  {journal}
  {\bibinfo  {journal} {Science}\ }\textbf {\bibinfo {volume} {365}},\ \bibinfo
  {pages} {1424} (\bibinfo {year} {2019})}\BibitemShut {NoStop}%
\bibitem [{\citenamefont {Jiang}\ \emph {et~al.}(2020)\citenamefont {Jiang},
  \citenamefont {Zaanen}, \citenamefont {Devereaux},\ and\ \citenamefont
  {Jiang}}]{Yifan2020}%
  \BibitemOpen
  \bibfield  {author} {\bibinfo {author} {\bibfnamefont {Y.-F.}\ \bibnamefont
  {Jiang}}, \bibinfo {author} {\bibfnamefont {J.}~\bibnamefont {Zaanen}},
  \bibinfo {author} {\bibfnamefont {T.~P.}\ \bibnamefont {Devereaux}},\ and\
  \bibinfo {author} {\bibfnamefont {H.-C.}\ \bibnamefont {Jiang}},\ }\bibfield
  {title} {\bibinfo {title} {Ground state phase diagram of the doped hubbard
  model on the four-leg cylinder},\ }\href
  {https://doi.org/10.1103/PhysRevResearch.2.033073} {\bibfield  {journal}
  {\bibinfo  {journal} {Phys. Rev. Res.}\ }\textbf {\bibinfo {volume} {2}},\
  \bibinfo {pages} {033073} (\bibinfo {year} {2020})}\BibitemShut {NoStop}%
\bibitem [{\citenamefont {Jiang}\ \emph {et~al.}(2021)\citenamefont {Jiang},
  \citenamefont {Scalapino},\ and\ \citenamefont {White}}]{Shengtao2021}%
  \BibitemOpen
  \bibfield  {author} {\bibinfo {author} {\bibfnamefont {S.}~\bibnamefont
  {Jiang}}, \bibinfo {author} {\bibfnamefont {D.~J.}\ \bibnamefont
  {Scalapino}},\ and\ \bibinfo {author} {\bibfnamefont {S.~R.}\ \bibnamefont
  {White}},\ }\bibfield  {title} {\bibinfo {title} {Ground state phase diagram
  of the {$t$-$t'$-$J$ Model}},\ }\href
  {https://doi.org/10.1073/pnas.2109978118} {\bibfield  {journal} {\bibinfo
  {journal} {Proc. Natl. Acad. Sci. U.S.A.}\ }\textbf {\bibinfo {volume}
  {118}},\ \bibinfo {pages} {e2109978118} (\bibinfo {year} {2021})}\BibitemShut
  {NoStop}%
\bibitem [{\citenamefont {Jiang}\ \emph
  {et~al.}(2023{\natexlab{b}})\citenamefont {Jiang}, \citenamefont
  {Devereaux},\ and\ \citenamefont {Jiang}}]{Jiang2023Six}%
  \BibitemOpen
  \bibfield  {author} {\bibinfo {author} {\bibfnamefont {Y.-F.}\ \bibnamefont
  {Jiang}}, \bibinfo {author} {\bibfnamefont {T.~P.}\ \bibnamefont
  {Devereaux}},\ and\ \bibinfo {author} {\bibfnamefont {H.-C.}\ \bibnamefont
  {Jiang}},\ }\href@noop {} {\bibinfo {title} {Ground state phase diagram and
  superconductivity of the doped hubbard model on six-leg square cylinders}}
  (\bibinfo {year} {2023}{\natexlab{b}}),\ \Eprint
  {https://arxiv.org/abs/2303.15541} {arXiv:2303.15541 [cond-mat.str-el]}
  \BibitemShut {NoStop}%
\bibitem [{\citenamefont {White}(1992)}]{White1992}%
  \BibitemOpen
  \bibfield  {author} {\bibinfo {author} {\bibfnamefont {S.~R.}\ \bibnamefont
  {White}},\ }\bibfield  {title} {\bibinfo {title} {Density matrix formulation
  for quantum renormalization groups},\ }\href
  {https://doi.org/10.1103/PhysRevLett.69.2863} {\bibfield  {journal} {\bibinfo
   {journal} {Phys. Rev. Lett.}\ }\textbf {\bibinfo {volume} {69}},\ \bibinfo
  {pages} {2863} (\bibinfo {year} {1992})}\BibitemShut {NoStop}%
\bibitem [{\citenamefont {White}(1993)}]{White1993}%
  \BibitemOpen
  \bibfield  {author} {\bibinfo {author} {\bibfnamefont {S.~R.}\ \bibnamefont
  {White}},\ }\bibfield  {title} {\bibinfo {title} {Density-matrix algorithms
  for quantum renormalization groups},\ }\href
  {https://doi.org/10.1103/PhysRevB.48.10345} {\bibfield  {journal} {\bibinfo
  {journal} {Phys. Rev. B}\ }\textbf {\bibinfo {volume} {48}},\ \bibinfo
  {pages} {10345} (\bibinfo {year} {1993})}\BibitemShut {NoStop}%
\bibitem [{\citenamefont {\"Ostlund}\ and\ \citenamefont
  {Rommer}(1995)}]{Ostlund1995}%
  \BibitemOpen
  \bibfield  {author} {\bibinfo {author} {\bibfnamefont {S.}~\bibnamefont
  {\"Ostlund}}\ and\ \bibinfo {author} {\bibfnamefont {S.}~\bibnamefont
  {Rommer}},\ }\bibfield  {title} {\bibinfo {title} {Thermodynamic limit of
  density matrix renormalization},\ }\href
  {https://doi.org/10.1103/PhysRevLett.75.3537} {\bibfield  {journal} {\bibinfo
   {journal} {Phys. Rev. Lett.}\ }\textbf {\bibinfo {volume} {75}},\ \bibinfo
  {pages} {3537} (\bibinfo {year} {1995})}\BibitemShut {NoStop}%
\bibitem [{\citenamefont {Fishman}\ \emph
  {et~al.}(2022{\natexlab{a}})\citenamefont {Fishman}, \citenamefont {White},\
  and\ \citenamefont {Stoudenmire}}]{Fishman2022}%
  \BibitemOpen
  \bibfield  {author} {\bibinfo {author} {\bibfnamefont {M.}~\bibnamefont
  {Fishman}}, \bibinfo {author} {\bibfnamefont {S.~R.}\ \bibnamefont {White}},\
  and\ \bibinfo {author} {\bibfnamefont {E.~M.}\ \bibnamefont {Stoudenmire}},\
  }\bibfield  {title} {\bibinfo {title} {{The ITensor Software Library for
  Tensor Network Calculations}},\ }\href
  {https://doi.org/10.21468/SciPostPhysCodeb.4} {\bibfield  {journal} {\bibinfo
   {journal} {SciPost Phys. Codebases}\ ,\ \bibinfo {pages} {4}} (\bibinfo
  {year} {2022}{\natexlab{a}})}\BibitemShut {NoStop}%
\bibitem [{\citenamefont {Fishman}\ \emph
  {et~al.}(2022{\natexlab{b}})\citenamefont {Fishman}, \citenamefont {White},\
  and\ \citenamefont {Stoudenmire}}]{Fishman2022_2}%
  \BibitemOpen
  \bibfield  {author} {\bibinfo {author} {\bibfnamefont {M.}~\bibnamefont
  {Fishman}}, \bibinfo {author} {\bibfnamefont {S.~R.}\ \bibnamefont {White}},\
  and\ \bibinfo {author} {\bibfnamefont {E.~M.}\ \bibnamefont {Stoudenmire}},\
  }\bibfield  {title} {\bibinfo {title} {{Codebase release 0.3 for ITensor}},\
  }\href {https://doi.org/10.21468/SciPostPhysCodeb.4-r0.3} {\bibfield
  {journal} {\bibinfo  {journal} {SciPost Phys. Codebases}\ ,\ \bibinfo {pages}
  {4}} (\bibinfo {year} {2022}{\natexlab{b}})}\BibitemShut {NoStop}%
\bibitem [{\citenamefont {Hybertsen}\ \emph {et~al.}(1989)\citenamefont
  {Hybertsen}, \citenamefont {Schl\"uter},\ and\ \citenamefont
  {Christensen}}]{Hybertsen1989}%
  \BibitemOpen
  \bibfield  {author} {\bibinfo {author} {\bibfnamefont {M.~S.}\ \bibnamefont
  {Hybertsen}}, \bibinfo {author} {\bibfnamefont {M.}~\bibnamefont
  {Schl\"uter}},\ and\ \bibinfo {author} {\bibfnamefont {N.~E.}\ \bibnamefont
  {Christensen}},\ }\bibfield  {title} {\bibinfo {title} {Calculation of
  coulomb-interaction parameters for ${\mathrm{la}}_{2}$${\mathrm{cuo}}_{4}$
  using a constrained-density-functional approach},\ }\href
  {https://doi.org/10.1103/PhysRevB.39.9028} {\bibfield  {journal} {\bibinfo
  {journal} {Phys. Rev. B}\ }\textbf {\bibinfo {volume} {39}},\ \bibinfo
  {pages} {9028} (\bibinfo {year} {1989})}\BibitemShut {NoStop}%
\bibitem [{\citenamefont {Nishimoto}\ \emph {et~al.}(2002)\citenamefont
  {Nishimoto}, \citenamefont {Jeckelmann},\ and\ \citenamefont
  {Scalapino}}]{Nishimoto2002}%
  \BibitemOpen
  \bibfield  {author} {\bibinfo {author} {\bibfnamefont {S.}~\bibnamefont
  {Nishimoto}}, \bibinfo {author} {\bibfnamefont {E.}~\bibnamefont
  {Jeckelmann}},\ and\ \bibinfo {author} {\bibfnamefont {D.~J.}\ \bibnamefont
  {Scalapino}},\ }\bibfield  {title} {\bibinfo {title} {Differences between
  hole and electron doping of a two-leg cuo ladder},\ }\href
  {https://doi.org/10.1103/PhysRevB.66.245109} {\bibfield  {journal} {\bibinfo
  {journal} {Phys. Rev. B}\ }\textbf {\bibinfo {volume} {66}},\ \bibinfo
  {pages} {245109} (\bibinfo {year} {2002})}\BibitemShut {NoStop}%
\bibitem [{\citenamefont {White}\ and\ \citenamefont
  {Scalapino}(2015)}]{White2015}%
  \BibitemOpen
  \bibfield  {author} {\bibinfo {author} {\bibfnamefont {S.~R.}\ \bibnamefont
  {White}}\ and\ \bibinfo {author} {\bibfnamefont {D.~J.}\ \bibnamefont
  {Scalapino}},\ }\bibfield  {title} {\bibinfo {title} {Doping asymmetry and
  striping in a three-orbital ${\mathrm{cuo}}_{2}$ hubbard model},\ }\href
  {https://doi.org/10.1103/PhysRevB.92.205112} {\bibfield  {journal} {\bibinfo
  {journal} {Phys. Rev. B}\ }\textbf {\bibinfo {volume} {92}},\ \bibinfo
  {pages} {205112} (\bibinfo {year} {2015})}\BibitemShut {NoStop}%
\bibitem [{\citenamefont {White}\ \emph {et~al.}(2002)\citenamefont {White},
  \citenamefont {Affleck},\ and\ \citenamefont {Scalapino}}]{White2002Friedel}%
  \BibitemOpen
  \bibfield  {author} {\bibinfo {author} {\bibfnamefont {S.~R.}\ \bibnamefont
  {White}}, \bibinfo {author} {\bibfnamefont {I.}~\bibnamefont {Affleck}},\
  and\ \bibinfo {author} {\bibfnamefont {D.~J.}\ \bibnamefont {Scalapino}},\
  }\bibfield  {title} {\bibinfo {title} {Friedel oscillations and charge
  density waves in chains and ladders},\ }\href
  {https://doi.org/10.1103/PhysRevB.65.165122} {\bibfield  {journal} {\bibinfo
  {journal} {Phys. Rev. B}\ }\textbf {\bibinfo {volume} {65}},\ \bibinfo
  {pages} {165122} (\bibinfo {year} {2002})}\BibitemShut {NoStop}%
\bibitem [{\citenamefont {Agterberg}\ and\ \citenamefont
  {Garaud}(2015)}]{Agterberg2015}%
  \BibitemOpen
  \bibfield  {author} {\bibinfo {author} {\bibfnamefont {D.~F.}\ \bibnamefont
  {Agterberg}}\ and\ \bibinfo {author} {\bibfnamefont {J.}~\bibnamefont
  {Garaud}},\ }\bibfield  {title} {\bibinfo {title} {Checkerboard order in
  vortex cores from pair-density-wave superconductivity},\ }\href
  {https://doi.org/10.1103/PhysRevB.91.104512} {\bibfield  {journal} {\bibinfo
  {journal} {Phys. Rev. B}\ }\textbf {\bibinfo {volume} {91}},\ \bibinfo
  {pages} {104512} (\bibinfo {year} {2015})}\BibitemShut {NoStop}%
\bibitem [{\citenamefont {Scalettar}\ \emph {et~al.}(1991)\citenamefont
  {Scalettar}, \citenamefont {Scalapino}, \citenamefont {Sugar},\ and\
  \citenamefont {White}}]{Scalettar1991}%
  \BibitemOpen
  \bibfield  {author} {\bibinfo {author} {\bibfnamefont {R.~T.}\ \bibnamefont
  {Scalettar}}, \bibinfo {author} {\bibfnamefont {D.~J.}\ \bibnamefont
  {Scalapino}}, \bibinfo {author} {\bibfnamefont {R.~L.}\ \bibnamefont
  {Sugar}},\ and\ \bibinfo {author} {\bibfnamefont {S.~R.}\ \bibnamefont
  {White}},\ }\bibfield  {title} {\bibinfo {title} {Antiferromagnetic,
  charge-transfer, and pairing correlations in the three-band hubbard model},\
  }\href {https://doi.org/10.1103/PhysRevB.44.770} {\bibfield  {journal}
  {\bibinfo  {journal} {Phys. Rev. B}\ }\textbf {\bibinfo {volume} {44}},\
  \bibinfo {pages} {770} (\bibinfo {year} {1991})}\BibitemShut {NoStop}%
\bibitem [{\citenamefont {Chen}\ \emph {et~al.}(2021)\citenamefont {Chen},
  \citenamefont {Wang}, \citenamefont {Rebec}, \citenamefont {Jia},
  \citenamefont {Hashimoto}, \citenamefont {Lu}, \citenamefont {Moritz},
  \citenamefont {Moore}, \citenamefont {Devereaux},\ and\ \citenamefont
  {Shen}}]{Chen2021}%
  \BibitemOpen
  \bibfield  {author} {\bibinfo {author} {\bibfnamefont {Z.}~\bibnamefont
  {Chen}}, \bibinfo {author} {\bibfnamefont {Y.}~\bibnamefont {Wang}}, \bibinfo
  {author} {\bibfnamefont {S.~N.}\ \bibnamefont {Rebec}}, \bibinfo {author}
  {\bibfnamefont {T.}~\bibnamefont {Jia}}, \bibinfo {author} {\bibfnamefont
  {M.}~\bibnamefont {Hashimoto}}, \bibinfo {author} {\bibfnamefont
  {D.}~\bibnamefont {Lu}}, \bibinfo {author} {\bibfnamefont {B.}~\bibnamefont
  {Moritz}}, \bibinfo {author} {\bibfnamefont {R.~G.}\ \bibnamefont {Moore}},
  \bibinfo {author} {\bibfnamefont {T.~P.}\ \bibnamefont {Devereaux}},\ and\
  \bibinfo {author} {\bibfnamefont {Z.-X.}\ \bibnamefont {Shen}},\ }\bibfield
  {title} {\bibinfo {title} {Anomalously strong near-neighbor attraction in
  doped {{1D}} cuprate chains},\ }\bibfield  {journal} {\bibinfo  {journal}
  {Science}\ }\href {https://doi.org/10.1126/science.abf5174}
  {10.1126/science.abf5174} (\bibinfo {year} {2021})\BibitemShut {NoStop}%
\bibitem [{\citenamefont {Qu}\ \emph {et~al.}(2022)\citenamefont {Qu},
  \citenamefont {Chen}, \citenamefont {Jiang}, \citenamefont {Wang},\ and\
  \citenamefont {Li}}]{Qu2022}%
  \BibitemOpen
  \bibfield  {author} {\bibinfo {author} {\bibfnamefont {D.-W.}\ \bibnamefont
  {Qu}}, \bibinfo {author} {\bibfnamefont {B.-B.}\ \bibnamefont {Chen}},
  \bibinfo {author} {\bibfnamefont {H.-C.}\ \bibnamefont {Jiang}}, \bibinfo
  {author} {\bibfnamefont {Y.}~\bibnamefont {Wang}},\ and\ \bibinfo {author}
  {\bibfnamefont {W.}~\bibnamefont {Li}},\ }\bibfield  {title} {\bibinfo
  {title} {Spin-triplet pairing induced by near-neighbor attraction in the
  extended hubbard model for cuprate chain},\ }\href
  {https://doi.org/10.1038/s42005-022-01030-x} {\bibfield  {journal} {\bibinfo
  {journal} {Commun Phys}\ }\textbf {\bibinfo {volume} {5}},\ \bibinfo {pages}
  {257} (\bibinfo {year} {2022})}\BibitemShut {NoStop}%
\bibitem [{\citenamefont {Peng}\ \emph {et~al.}(2023)\citenamefont {Peng},
  \citenamefont {Wang}, \citenamefont {Wen}, \citenamefont {Lee}, \citenamefont
  {Devereaux},\ and\ \citenamefont {Jiang}}]{Peng2023}%
  \BibitemOpen
  \bibfield  {author} {\bibinfo {author} {\bibfnamefont {C.}~\bibnamefont
  {Peng}}, \bibinfo {author} {\bibfnamefont {Y.}~\bibnamefont {Wang}}, \bibinfo
  {author} {\bibfnamefont {J.}~\bibnamefont {Wen}}, \bibinfo {author}
  {\bibfnamefont {Y.~S.}\ \bibnamefont {Lee}}, \bibinfo {author} {\bibfnamefont
  {T.~P.}\ \bibnamefont {Devereaux}},\ and\ \bibinfo {author} {\bibfnamefont
  {H.-C.}\ \bibnamefont {Jiang}},\ }\bibfield  {title} {\bibinfo {title}
  {Enhanced superconductivity by near-neighbor attraction in the doped extended
  hubbard model},\ }\href {https://doi.org/10.1103/PhysRevB.107.L201102}
  {\bibfield  {journal} {\bibinfo  {journal} {Phys. Rev. B}\ }\textbf {\bibinfo
  {volume} {107}},\ \bibinfo {pages} {L201102} (\bibinfo {year}
  {2023})}\BibitemShut {NoStop}%
\end{thebibliography}%
\end{document}